\documentclass[11pt,oneside]{article}%
\usepackage{amsmath}
\usepackage{amssymb}
\usepackage{amsfonts}
\usepackage{graphicx}
\usepackage{cite}%
\allowdisplaybreaks
\newcommand{\eqa}{\begin{eqnarray}}
\newcommand{\ena}{\end{eqnarray}}

\begin{document}

\author{Alexander Schmidt\thanks{e-mail: schmidt@theorie.physik.uni-muenchen.de},
Hartmut Wachter\thanks{e-mail: Hartmut.Wachter@physik.uni-muenchen.de}%
\vspace{0.2cm}\vspace{0.4cm}\\Max-Planck-Institute\\for Mathematics in the Sciences \\Inselstr. 22, D-04103 Leipzig, Germany\\\hspace{0.4in}\\Arnold-Sommerfeld-Center\\Ludwig-Maximilians-Universit\"{a}t\\Theresienstr. 37, D-80333 M\"{u}nchen, Germany}
\title{Superanalysis on quantum spaces}
\date{}
\maketitle

\begin{abstract}
\noindent Attention is focused on antisymmetrised versions of quantum spaces
that are of particular importance in physics, i.e. Manin plane, q-deformed
Euclidean space in three or four dimensions as well as q-deformed Minkowski
space. For each of these quantum spaces we provide q-analogs for elements of
superanalysis, i.e. Grassmann integrals, Grassmann exponentials, Grassmann
translations and braided products with supernumbers.\vspace{0.5cm}\newline
\newpage

\end{abstract}

\section{Introduction}

\noindent\textbf{General motivation for deforming spacetime}

Relativistic quantum field theory is not a fundamental theory, since its
formalism leads to divergencies. In some cases like that of quantum
electrodynamics one is able to overcome the difficulties with the divergencies
by applying the so-called renormalization procedure due to Richard Feynman.
Unfortunately, this procedure is not successful if we want to deal with
quantum gravity. Despite the fact that gravitation is a rather weak
interaction we are not able to treat it perturbatively. The reason for this
lies in the fact, that transition amplitudes of nth order to the gravitation
constant diverge like a momentum integral of the general form \cite{Wei}
\begin{equation}
\int p^{2n-1}dp, \label{ImInt}%
\end{equation}
leaving us with an infinite number of ultraviolet divergent Feynman diagrams
that cannot be removed by redefining finitely many physical parameters.

It is surely legitimate to ask for the reason for these fundamental
difficulties. It is commonplace that the problems with the divergences in
relativistic quantum field theory result from an incomplete description of
spacetime at very small distances \cite{Schw}. Niels Bohr and Werner
Heisenberg have been the first who suggested that quantum field theories
should be formulated on a spacetime lattice \cite{Cass, Heis}. Such a
spacetime lattice would imply the existence of a smallest distance $a$ with
the consequence that plane-waves of wave-length smaller than twice the lattice
spacing could not propagate. In accordance with the relationship between
wave-length $\lambda$ and momentum $p$ of a plane-wave, i.e.
\begin{equation}
\lambda\geq\lambda_{\min}=2a\quad\Rightarrow\quad\frac{1}{\lambda}\sim p\leq
p_{\max}\sim\frac{1}{2a},
\end{equation}
it follows then that physical momentum space would be bounded. Hence, the
domain of all momentum integrals in Eq. (\ref{ImInt}) would be bounded as well
with the consequence that momentum integrals should take on finite
values.\textsf{\\[0.05in]}

\noindent\textbf{q-Deformation of symmetries as an attempt to get a more
detailed description of nature}

Discrete spacetime structures in general do not respect classical Poincar\'{e}
symmetry. A possible way out of this difficulty is to modify not only
spacetime but also its corresponding symmetries. How are we to accomplish
this? First of all let us recall that classical spacetime symmetries are
usually described by Lie groups. Realizing that Lie groups are manifolds the
Gelfand-Naimark\textsf{\ }theorem tells us that Lie groups can be naturally
embedded in the category of algebras\textsf{\ }\cite{GeNe}. The utility of
this interrelation lies in formulating the geometrical structure of Lie groups
in terms of a Hopf structure \cite{Hopf}. The point is that during the last
two decades generic methods have been discovered for continuously deforming
matrix groups and Lie algebras within the category of Hopf algebras. It is
this development which finally led to the arrival of quantum groups and
quantum spaces \cite{Ku83, Wor87, Dri85, Jim85, Drin86, RFT90, Tak90}.

From a physical point of view the most realistic and interesting deformations
are given by q-deformed versions of Minkowski space and Euclidean spaces as
well as their corresponding symmetries, i.e. respectively Lorentz symmetry and
rotational symmetry \cite{CSSW90, Pod90, SWZ91, Maj91, LWW97}. Further studies
even allowed to establish differential calculi on these q-deformed quantum
spaces \cite{WZ91, CSW91, OSWZ92} representing nothing other than q-analogs of
classical translational symmetry. In this sense we can say that q-deformations
of the complete Euclidean and Poincar\'{e} symmetries are now available
\cite{Maj-93/1}. Finally, Julius Wess and his coworkers were able to show that
q-deformation of spaces and symmetries can indeed leed to the wanted
discretizations of the spectra of spacetime observables \cite{Fich97, CW98},
which nourishes the hope that q-deformation might give a new method to
regularize quantum field theories \cite{GKP96, MajReg, Oec99, Blo03}%
.\textsf{\\[0.05in]}

\noindent\textbf{Foundations of q-deformed superanalysis}

In order to formulate quantum field theories on q-deformed quantum spaces it
is necessary to provide us with some essential tools of a q-deformed analysis.
The main question is how to define these new tools, which should be q-analogs
of classical notions. Towards this end the considerations of Shahn Majid have
proved very useful \cite{Maj91Kat, Maj94Kat, Maj94-10}. The key idea of his
approach is that all the quantum spaces to a given quantum symmetry form a
braided tensor category. Consequently, operations and objects concerning
quantum spaces must rely on this framework of a braided tensor category, in
order to guarantee their well-defined behavior under quantum group
transformations. This so-called principle of covariance can be seen as the
essential guideline for constructing a consistent theory.

In our previous work we have worked on symmetrized versions of quantum spaces
that are of particular importance in physics, i.e. Manin plane, q-deformed
Euclidean space with three or four dimensions as well as q-deformed Minkowski
space. In each case we have presented explicit formulae for star-products
\cite{WW01}, representations of symmetry generators and partial derivatives
\cite{BW01}, q-integrals \cite{Wac02}, q-exponentials \cite{Wac04} and
q-translations \cite{Wac04-10}. But physics requires also antisymmetrized
spaces, i.e. Grassmann algebras, since they constitute an important tool in
formulating supersymmetrical quantum field theories. In Ref. \cite{Mik04} we
started showing that our ideas for symmetrized quantum spaces carry over to
antisymmetrized ones as well.

Our goal now is to continue that program by providing explicit formulae for
q-analogs of Grassmann integrals, Grassmann exponentials and Grassmann
translations. In addition to this we are going to present formulae for braided
products with supernumbers telling us how antisymmetrized quantum spaces have
to be fused together with other quantum spaces.

The paper is organized as follows. In Sec. \ref{Ideas} we give a review of the
concepts q-deformed superanalysis is based on. Furthermore, we recall some
important results of Ref. \cite{Mik04}. This is done to an extent necessary
for our further studies. In Sec. \ref{QuanPlan} we explain in detail how our
general considerations apply to Manin plane. In Secs. \ref{Sec3} and
\ref{Sec4} we repeat the same steps as in Sec. \ref{QuanPlan} for q-deformed
Euclidean spaces in three and four dimensions, respectively. Sec.
\ref{MinSpac} is devoted to superanalysis on an antisymmetrized version of
q-deformed Minkowski. Finally, in Sec. \ref{AppA} we give a short conclusion
und provide the reader with some interesting remarks about our new objects.

\section{Concepts of q-deformed superanalysis\label{Ideas}}

As already mentioned, q-deformed superanalysis is formulated within the
framework of antisymmetrized quantum spaces. These quantum spaces are defined
as modules of quasitriangular Hopf algebras which describe the underlying
symmetry. For our purposes, it is at first sufficient to consider an
antisymmetrized quantum space as an algebra generated by coordinates
$\theta^{1},\theta^{2},\ldots,\theta^{n}$ which are subjected to
\begin{equation}
\theta^{i}\theta^{j}=-k(\hat{R})^{ij}{}_{kl}\theta^{k}\theta^{l},\quad
k\in\mathbb{R}^{+}, \label{RelGrass2dim}%
\end{equation}
where $\hat{R}$ denotes a representation of the universal $\mathcal{R}$-matrix
assigned to\ the underlying quantum symmetry. This way, we get nothing other
than q-deformed versions of Grassman algebras.

Moreover, it is important to realize that our antisymmetrized quantum spaces
satisfy the so-called \textit{Poincar\'{e}-Birkhoff-Witt property}, i.e. the
dimension of a subspace of homogenous polynomials should be the same as for
classical Grassmann variables. This property is the deeper reason why normal
ordered monomials constitute a basis of our q-deformed Grassmann algebras.
Consequently, each supernumber can be represented in the general form
\begin{equation}
f(\underline{\theta})=f^{\prime}+\sum\nolimits_{\underline{K}}f_{\underline
{K}}\,\theta^{\,\underline{K}}, \label{SupNumAllg}%
\end{equation}
where $\theta^{\underline{K}}$ denotes monomials of a given normal ordering.

For this to become more clear, the two-dimensional antisymmetrized quantum
plane shall serve as an example \cite{Man88}. By specifying the R-matrix in
Eq. (\ref{RelGrass2dim})\ to that of $U_{q}(su_{2}),$ we obtain as defining
relations of antisymmetrized Manin plane%
\begin{align}
(\theta^{1})^{2}  &  =(\theta^{2})^{2}=0,\\
\theta^{1}\theta^{2}  &  =-q^{-1}\theta^{2}\theta^{1},\nonumber
\end{align}
showing the correct classical limit for $q\rightarrow1.$ Due to these
relations, each supernumber can be written in the general form (notice that
the normal ordering of monomials is indicated by the order in which
coordinates are arranged in the symbol for supernumbers)%
\begin{equation}
f(\theta^{2},\theta^{1})=f^{\prime}+f_{1}\theta^{1}+f_{2}\theta^{2}%
+f_{21}\theta^{2}\theta^{1}, \label{RepSup2dim}%
\end{equation}
and the product of two such supernumbers finally becomes
\begin{align}
&  (f\cdot g)(\theta^{2},\theta^{1})\\
&  =(f\cdot g)^{\prime}+(f\cdot g)_{1}\theta^{1}+(f\cdot g)_{2}\theta
^{2}+(f\cdot g)_{21}\theta^{2}\theta^{1},\nonumber
\end{align}
with
\begin{align}
(f\cdot g)^{\prime}  &  =f^{\prime}g^{\prime},\\
(f\cdot g)_{\alpha}  &  =f_{\alpha}g^{\prime}+f^{\prime}g_{\alpha},\quad
\alpha=1,2,\nonumber\\
(f\cdot g)_{21}  &  =f_{2}g_{1}-q^{-1}f_{1}g_{2}.\nonumber
\end{align}
Similar results hold for the other antisymmetrized quantum spaces we consider
in this article \cite{Mik04}.

Next, we would like to come to the covariant differential calculi on our
antisymmetrized quantum spaces \cite{WZ91, CSW91, Song92}. In complete
accordance to symmetrized quantum spaces, there exist always two covariant
differential calculi. Their Leibniz rules take\ the general form%
\begin{align}
\partial_{\theta}^{i}\theta^{j}  &  =g^{ij}-k(\hat{R}^{-1})^{ij}{}%
_{kl}\,\theta^{k}\partial_{\theta}^{l},\quad k\in\mathbb{C},\\
\hat{\partial}_{\theta}^{i}\theta^{j}  &  =g^{ij}-k^{-1}(\hat{R})^{ij}{}%
_{kl}\,\theta^{k}\hat{\partial}_{\theta}^{l},\nonumber
\end{align}
where $g^{ij}$ denotes the in the corresponding quantum metric (as reference,
we provide a review of key notations in Appendix \ref{AppQua}). In the
two-dimensional case, for example, the relations for the first differential
calculus read explicitly%
\begin{align}
\partial_{\theta}^{1}\theta^{1}  &  =-q^{-1}\theta^{1}\partial_{\theta}%
^{1},\label{VerPartKoord2dim}\\
\partial_{\theta}^{1}\theta^{2}  &  =-q^{-1/2}-\theta^{2}\partial_{\theta}%
^{1},\nonumber\\[0.16in]
\partial_{\theta}^{2}\theta^{1}  &  =q^{1/2}-\theta^{1}\partial_{\theta}%
^{2}+\lambda\theta^{2}\partial_{\theta}^{1},\\
\partial_{\theta}^{2}\theta^{2}  &  =-q^{-1}\theta^{2}\partial_{\theta}%
^{2},\nonumber
\end{align}
leading to the following actions on supernumbers \cite{Mik04}:%
\begin{align}
(\partial_{\theta})_{1}\rhd f(\theta^{2},\theta^{1})  &  =f_{1}-q^{-1}%
f_{21}\theta^{2},\label{PartDef2dim}\\
(\partial_{\theta})_{2}\rhd f(\theta^{2},\theta^{1})  &  =f_{2}+f_{21}%
\theta^{1}.\nonumber
\end{align}

However, in what follows it is necessary to take another point of view which
is provided by category theory. A category is a collection of objects
$X,Y,Z,\ldots$ together with a set Mor$(X,Y)$ of morphisms between two objects
$X,Y$. The composition of morphisms has similar properties as the composition
of maps. We are interested in tensor categories. These categories have a
product, denoted $\otimes$ and called the tensor product. It admits several
'natural' properties such as associativity and existence of a unit object. For
a more formal treatment we refer to Refs. \cite{Maj91Kat, Maj94Kat},
\cite{Maj95}or \cite{MaL74}. If the action of a quasitriangular Hopf algebra
$\mathcal{H}$ on the tensor product of two quantum spaces $X$ and $Y$ is
defined by
\begin{equation}
h\triangleright(v\otimes w)=(h_{(1)}\triangleright v)\otimes(h_{(2)}%
\triangleright w)\in X\otimes Y,\quad h\in\mathcal{H},
\end{equation}
where the coproduct is written in the so-called Sweedler notation, i.e.
$\Delta(h)=h_{(1)}\otimes h_{(2)},$ then the representations (quantum spaces)
of the given Hopf algebra (quantum algebra) are the objects of a tensor category.

In this tensor category exist a number of morphisms of particular importance
that are covariant with respect to the Hopf algebra action. First of all, for
any pair of objects $X,Y$ there is an isomorphism $\Psi_{X,Y}:X\otimes
Y\rightarrow Y\otimes X$ such that $(g\otimes f)\circ\Psi_{X,Y}=\Psi
_{X^{\prime},Y^{\prime}}\circ(f\otimes g)$ for arbitrary morphisms $f\in$
Mor$(X,X^{\prime})$ and $g\in$ Mor$(Y,Y^{\prime})$. In addition to this one
requires the hexagon axiom to hold. The hexagon axiom is the validity of the
two conditions
\begin{equation}
\Psi_{X,Y}\circ\Psi_{Y,Z}=\Psi_{X\otimes Y,Z},\quad\Psi_{X,Z}\circ\Psi
_{X,Y}=\Psi_{X,Y\otimes Z}.
\end{equation}
A tensor category equipped with such mappings $\Psi_{X,Y}$ for each pair of
objects $X,Y$ is called a braided tensor category. The mappings $\Psi_{X,Y}$
as a whole are often referred to as the braiding of the tensor category.
Furthermore, for any quantum space algebra $X$ in this category there are
morphisms $\Delta:X\rightarrow X\otimes X,$ $S:X\rightarrow X$ and
$\varepsilon:X\rightarrow\mathbb{C}$ forming a braided Hopf algebra, i.e.
$\Delta,$ $S$ and $\epsilon$ obey the usual axioms of a Hopf algebra, but now
as morphisms in the braided category. For further details we recommend
Refs.\ \cite{Maj95} and \cite{ChDe96}.

The explicit form of these morphisms is completely determined by the so-called
$\mathcal{L}$-matrix \cite{RFT90, Maj-93/1, Tan3}. The entries of the
$\mathcal{L}$-matrix are built up out of symmetry generators and scaling
operators. For the quantum spaces we study in this article, the explicit form
of the $\mathcal{L}$-matrix can be looked up in Ref. \cite{Mik04}. To be more
concrete, we give as example the non-vanishing entries of the $\mathcal{L}%
$-matrix and its conjugate in the case of Manin plane:%
\begin{align}
(\mathcal{L}_{a})_{1}^{1}  &  =\Lambda(a)\tau^{-1/4},\\
(\mathcal{L}_{a})_{1}^{2}  &  =-q\lambda\Lambda(a)\tau^{-1/4}T^{+},\nonumber\\
(\mathcal{L}_{a})_{2}^{2}  &  =\Lambda(a)\tau^{1/4},\nonumber
\end{align}
and likewise%
\begin{align}
(\mathcal{\bar{L}}_{a})_{1}^{1}  &  =\Lambda^{-1}(a)\tau^{1/4},\\
(\mathcal{\bar{L}}_{a})_{2}^{1}  &  =-q^{-1}\lambda\Lambda^{-1}(a)\tau
^{-1/4}T^{-},\nonumber\\
(\mathcal{\bar{L}}_{a})_{2}^{2}  &  =\Lambda^{-1}(a)\tau^{1/4},\nonumber
\end{align}
where $\tau^{\pm1/4}$, $T^{\pm}$ and $\Lambda(a)$ denote generators of the
quantum algebra $U_{q}(su_{2})$ and a unitary scaling operator, respectively.

Using the $\mathcal{L}$-matrix and its conjugate, the two distinct braidings
of a quantum space generator $a^{i}$ can be obtained in the compact form
\begin{align}
\Psi_{X,Y}(a^{i}\otimes f)  &  =((\mathcal{\bar{L}}_{a})_{j}^{i}\triangleright
f)\otimes a^{j},\label{BraidAlgGen}\\
\Psi_{X,Y}^{-1}(a^{i}\otimes f)  &  =((\mathcal{L}_{a})_{j}^{i})\triangleright
f)\otimes a^{j}\nonumber\\[0.16in]
\Psi_{X,Y}(f\otimes a^{i})  &  =a^{j}\otimes(f\triangleleft(\mathcal{L}%
_{a})_{j}^{i}),\\
\Psi_{X,Y}^{-1}(f\otimes a^{i})  &  =a^{j}\otimes(f\triangleleft
(\mathcal{\bar{L}}_{a})_{j}^{i}).\nonumber
\end{align}
There remains to evaluate the action of the $\mathcal{L}$-matrix on the
quantum space element $f.$ This can be achieved by making use of the
representations presented in Refs. \cite{BW01} and \cite{Mik04}.

By repeated use of the identites in Eq. (\ref{BraidAlgGen}) we are able to
calculate the braiding between a monomial in Grassmann variables $\theta^{i}%
$\ and an arbitrary element $g$ of another quantum space, i.e.%
\begin{align}
(\theta^{i}\ldots\theta^{j})\,\underline{\odot}_{\bar{L}}\,g  &  \equiv
\Psi(\theta^{i}\ldots\theta^{j}\otimes g)\label{BraiBasTheta}\\
&  =\left(  (\mathcal{\bar{L}}_{\theta})_{k_{i}}^{i}\ldots(\mathcal{\bar{L}%
}_{\theta})_{k_{j}}^{j}\rhd g\right)  \otimes(\theta^{k_{i}}\ldots
\theta^{k_{j}}),\nonumber\\
(\theta^{i}\ldots\theta^{j})\,\underline{\odot}_{{L}}\,g  &  \equiv\Psi
^{-1}(\theta^{i}\ldots\theta^{j}\otimes g)\nonumber\\
&  =\left(  (\mathcal{L}_{\theta})_{k_{i}}^{i}\ldots(\mathcal{L}_{\theta
})_{k_{j}}^{j}\rhd g\right)  \;\otimes(\theta^{k_{i}}\ldots\theta^{k_{j}%
}),\nonumber\\%
[0.1in]%
g\;\underline{\odot}_{R}\,(\theta^{i}\ldots\theta^{j})  &  \equiv\Psi
(g\otimes\theta^{i}\ldots\theta^{j})\\
&  =(\theta^{k_{i}}\ldots\theta^{k_{j}})\otimes\left(  g\lhd(\mathcal{L}%
_{\theta})_{k_{i}}^{i}\ldots(\mathcal{L}_{\theta})_{k_{j}}^{j}\right)
,\nonumber\\
g\;\underline{\odot}_{\bar{R}}\,(\theta^{i}\ldots\theta^{j})  &  \equiv
\Psi^{-1}(g\otimes\theta^{i}\ldots\theta^{j})\nonumber\\
&  =(\theta^{k_{i}}\ldots\theta^{k_{j}})\otimes\left(  g\lhd(\mathcal{\bar{L}%
}_{\theta})_{k_{i}}^{i}\ldots(\mathcal{\bar{L}}_{\theta})_{k_{j}}^{j}\right)
.\nonumber
\end{align}
Recalling that the braiding mappings are linear in their arguments, it should
be quite clear that the braiding of a supernumber with an arbitrary element
$g$ is completely determined by the above identities. In the subsequent
sections this observation will enable us to derive explicit formulae for the
braiding of supernumbers with arbitrary quantum space elements. The resulting
expressions are referred to as braided products for supernumbers.

Using $\mathcal{L}$-matrices, the coproducts for quantum space generators
$a^{i}$ can be obtained in the general form
\begin{align}
\Delta_{\bar{L}}(a^{i})  &  =a^{i}\otimes1+(\mathcal{\bar{L}}_{a})_{j}%
^{i}\otimes a^{j},\label{CoPrGenForm}\\
\Delta_{L}(a^{i})  &  =a^{i}\otimes1+(\mathcal{L}_{a})_{j}^{i}\otimes
a^{j},\nonumber
\end{align}
and the corresponding antipodes are then given by (if we assume for the counit
$\varepsilon(\theta^{\iota})=0$)%
\begin{align}
S_{\bar{L}}(\theta^{i})  &  =-S(\mathcal{\bar{L}}_{\theta})_{j}^{i}%
\,\theta^{j},\label{AntGen}\\
S_{L}(\theta^{i})  &  =-S(\mathcal{L}_{\theta})_{j}^{i}\,\theta^{j}.\nonumber
\end{align}
The essential observation for this paper is that coproducts of coordinates
imply their translations \cite{Maj-93/1, Maj-93/2, Maj-93/3, Me95}. This can
be seen as follows. The coproduct $\Delta$ on coordinates is an algebra
homomorphism. If the coordinates constitute a module coalgebra then the
algebra structure of the coordinates $a^{i}$ is carried over to their
coproduct $\Delta(a^{i}).$ More formally, we have
\begin{equation}
\Delta(a^{i}a^{j})=\Delta(a^{i})\Delta(a^{j})\quad\text{and\quad}\Delta(h\rhd
a^{i})=\Delta(h)\rhd\Delta(a^{i}).
\end{equation}
Due to this fact we can think of (\ref{CoPrGenForm}) as nothing other than an
addition law for q-deformed vector components. In this article we use this
fact to derive translation operations for q-deformed supernumbers.

Next, let us make contact with another important ingredient of q-deformed
superanalysis, i.e. q-deformed Grassmann exponential. For this purpose we have
to suppose that our category is equipped with a dual object $X^{\ast}$ for
each algebra $X$ in the category. This means that we have dual pairings
\begin{equation}
\langle\;,\;\rangle:X\otimes X^{\ast}\rightarrow\mathbb{C}\quad
\mbox{with}\quad\langle e_{a},f^{b}\rangle=\delta_{a}^{b}, \label{DualParr}%
\end{equation}
where ${\{}e_{a}{\}}$ is a basis in $X$ and ${\{}f_{a}{\}}$ a dual basis in
$X^{\ast}$. Now, we are able to introduce an exponential map \cite{Maj-93/2}
which is defined to be the dual object of (\ref{DualParr}). Thus, the
exponential is given by
\begin{equation}
\mbox{exp}:\mathbb{C}\rightarrow X^{\ast}\otimes X,\quad\mbox{with}\quad
\mbox{ exp}=\sum_{a}f^{a}\otimes e_{a}. \label{eform}%
\end{equation}

It was shown in Ref. \cite{Maj94-10} that there is such a dual pairing of
Grassmann variables and corresponding partial derivatives. Specifically, we
have
\begin{equation}
\langle\;,\;\rangle:\mathcal{M}_{\partial}\otimes\mathcal{M}_{\theta
}\rightarrow\mathbb{C}\quad\mbox{with}\quad\langle f(\underline{\partial
}_{\theta}),g(\underline{\theta})\rangle\equiv\varepsilon(f(\underline
{\partial}_{\theta})\rhd g(\underline{\theta})).
\end{equation}
If we know a basis of the coordinate algebra $\mathcal{M}_{\theta}$ being dual
to a given one of $\mathcal{M}_{\partial}$, then we will be able to read off
from the definition in Eq. (\ref{eform}) the explicit form of the
q-exponentials. This task will be done in the subsequent sections for all
quantum spaces under consideration.

\section{Two-Dimensional quantum plane\label{QuanPlan}}

In this section we present explicit formulae for elements of q-deformed
superanalysis on antisymmetrized Manin plane (for its definition see also
Appendix \ref{AppQua}). To begin with, we introduce the notion of a
left-superintegral. With the partial derivatives obeying the relations in Eq.
(\ref{VerPartKoord2dim}) this can be done in complete analogy to the
undeformed case:
\begin{equation}
\int f(\theta^{2},\theta^{1})\;d_{L}^{2}\theta\equiv(\partial_{\theta}%
)_{1}(\partial_{\theta})_{2}\rhd f(\theta^{2},\theta^{1})=f_{21}.
\label{SupInt2dim}%
\end{equation}
This integral shows the same properties as its classical counterpart. Thus, it
is linear, normed and translationally invariant. Linearity is clear because of
the linearity of the derivatives and the other two properties follow from its
very definition, since we have
\begin{equation}
\int\theta^{2}\theta^{1}\,d_{L}^{2}\theta=1,\quad\int\theta^{\alpha}%
\,d_{L}^{2}\theta=\int d_{L}^{2}\theta=0,\quad\alpha=1,2,
\end{equation}
and
\begin{equation}
\int(\partial_{\theta})_{\alpha}\rhd f(\theta^{2},\theta^{1})\;d_{L}%
^{2}{\theta}=0,\quad\alpha=1,2.
\end{equation}

We could also have started our considerations with the conjugated partial
derivatives $(\hat{\partial}_{\theta})_{\alpha}$ whose representations are
linked to those in (\ref{PartDef2dim}) via the correspondence (see Ref.
\cite{Mik04})%
\begin{equation}
(\hat{\partial}_{\theta})_{\alpha}\,\bar{\rhd}\,f(\theta^{1},\theta
^{2})\overset{%
\genfrac{}{}{0pt}{}{\alpha\leftrightarrow\alpha^{\prime}}{q\leftrightarrow1/q}%
}{\longleftrightarrow}-(\partial_{\theta})_{\alpha^{\prime}}\rhd f(\theta
^{2},\theta^{1}),
\end{equation}
where $\alpha^{\prime}=3-\alpha$. The symbol $\overset{%
\genfrac{}{}{0pt}{}{\alpha\leftrightarrow\alpha^{\prime}}{q\leftrightarrow1/q}%
}{\longleftrightarrow}$ denotes a transition between the two expressions via
the substitutions%
\begin{align}
&  \theta^{\alpha}\overset{%
\genfrac{}{}{0pt}{}{\alpha\leftrightarrow\alpha^{\prime}}{q\leftrightarrow1/q}%
}{\longleftrightarrow}\theta^{\alpha^{\prime}},\quad\theta^{\alpha}%
\theta^{\beta}\overset{%
\genfrac{}{}{0pt}{}{\alpha\leftrightarrow\alpha^{\prime}}{q\leftrightarrow1/q}%
}{\longleftrightarrow}\theta^{\alpha^{\prime}}\theta^{\beta^{\prime}},\quad
q\overset{%
\genfrac{}{}{0pt}{}{\alpha\leftrightarrow\alpha^{\prime}}{q\leftrightarrow1/q}%
}{\longleftrightarrow}q^{-1},\\
&  f^{^{\prime}}\overset{%
\genfrac{}{}{0pt}{}{\alpha\leftrightarrow\alpha^{\prime}}{q\leftrightarrow1/q}%
}{\longleftrightarrow}f^{\prime},\quad f_{\alpha}\overset{%
\genfrac{}{}{0pt}{}{\alpha\leftrightarrow\alpha^{\prime}}{q\leftrightarrow1/q}%
}{\longleftrightarrow}f_{\alpha^{\prime}},\quad f_{\alpha\beta}\overset{%
\genfrac{}{}{0pt}{}{\alpha\leftrightarrow\alpha^{\prime}}{q\leftrightarrow1/q}%
}{\longleftrightarrow}f_{\alpha^{\prime}\beta^{\prime}},\quad\alpha
,\beta=1,2.\nonumber
\end{align}
The corresponding superintegral then becomes
\begin{equation}
\int f({\theta}^{1},\theta^{2})\;d_{\bar{L}}^{2}{\theta}=(\hat{\partial
}_{\theta})_{2}(\hat{\partial}_{\theta})_{1}\,\bar{\rhd}\,f({\theta}%
^{1},\theta^{2})=f_{12}. \label{SupInCon2dim}%
\end{equation}
Notice that in Eqs. (\ref{SupInt2dim}) and (\ref{SupInCon2dim}) the subscripts
at the integration measure help us to distinguish the two types of
superintegrals. Using the action of conjugated partial derivatives on
Grassmann variables, it is again straightforward to show that
\begin{equation}
\int\theta^{1}\theta^{2}\,d_{\bar{L}}^{2}\theta=1,\quad\int\theta^{\alpha
}\,d_{\bar{L}}^{2}\theta=\int d_{\bar{L}}^{2}\theta=0,\quad\alpha=1,2,
\end{equation}
and
\begin{equation}
\int(\partial_{\theta})_{\alpha}\,\bar{\rhd}\,f({\theta}^{1},\theta
^{2})\;d_{\bar{L}}^{2}\theta=0,\quad\alpha=1,2.
\end{equation}

However, superintegrals can also be constructed from partial derivatives
acting on a supernumber from the right. Let us recall that left and right
derivatives are related to each other by (see Ref. \cite{Mik04})
\begin{align}
&  f(\theta^{1},\theta^{2})\lhd(\hat{\partial}_{\theta})_{\alpha}%
\overset{\alpha\leftrightarrow\alpha^{\prime}}{\longleftrightarrow}%
-(\hat{\partial_{\theta}})_{\alpha^{\prime}}\,\bar{\rhd}\,f(\theta^{1}%
,\theta^{2}),\\
&  f(\theta^{2},\theta^{1})\,\bar{\lhd}\,(\partial_{\theta})_{\alpha}%
\overset{\alpha\leftrightarrow\alpha^{\prime}}{\longleftrightarrow}%
-(\partial_{\theta})_{\alpha^{\prime}}\rhd f(\theta^{2},\theta^{1}),\nonumber
\end{align}
where $\overset{\alpha\leftrightarrow\alpha^{\prime}}{\longleftrightarrow}$
now stands for the transition
\begin{align}
&  \theta^{\alpha}\overset{\alpha\leftrightarrow\alpha^{\prime}}%
{\longleftrightarrow}\theta^{\alpha\prime},\quad\theta^{\alpha}\theta^{\beta
}\overset{\alpha\leftrightarrow\alpha^{\prime}}{\longleftrightarrow}%
\theta^{\beta^{\prime}}\theta^{\alpha^{\prime}},\\
&  f^{\prime}\overset{\alpha\leftrightarrow\alpha^{\prime}}%
{\longleftrightarrow}f^{\prime},\quad f^{\alpha}\overset{\alpha\leftrightarrow
\alpha^{\prime}}{\longleftrightarrow}f^{\alpha\prime},\quad f^{\alpha}%
{}^{\beta}\overset{\alpha\leftrightarrow\alpha^{\prime}}{\longleftrightarrow
}f^{\beta^{\prime}\alpha^{\prime}}.\nonumber
\end{align}
Using right derivatives, we can proceed in very much the same way as was done
for left derivatives. This way we introduce
\begin{align}
&  \int d_{R}^{2}{\theta}\;f(\theta^{1},\theta^{2})\equiv f(\theta^{1}%
,\theta^{2})\lhd(\hat{\partial}_{\theta})_{2}(\hat{\partial}_{\theta}%
)_{1}=f_{12},\\
&  \int d_{\bar{R}}^{2}{\theta}\;f(\theta^{2},\theta^{1})=f(\theta^{2}%
,\theta^{1})\,\bar{\lhd}\,(\partial_{\theta})_{1}(\partial_{\theta}%
)_{2}=f_{21}.\nonumber
\end{align}
The new definitions lead immediately to
\begin{align}
&  \int d_{R}^{2}\theta\;\theta^{1}\theta^{2}=1,\quad\int d_{R}^{2}%
\theta\;\theta^{\alpha}=\int d_{R}^{2}\theta=0,\\
&  \int d_{\bar{R}}^{2}\theta\;{\theta}^{2}{\theta}^{1}=1,\quad\int d_{\bar
{R}}^{2}{\theta}\;\theta^{\alpha}=\int d_{\bar{R}}^{2}{\theta}=0,\quad{\alpha
}=1,2,\nonumber
\end{align}
and
\begin{align}
&  \int d_{R}^{2}{\theta}\;f(\theta^{1},\theta^{2})\lhd\hat{\partial}_{\theta
}^{\alpha}=0,\\
&  \int d_{\bar{R}}^{2}\theta\;f(\theta^{2},\theta^{1})\;\bar{\lhd}%
\;{\partial}^{\alpha}=0,\quad\alpha=1,2.\nonumber
\end{align}

Next, we come to the q-deformed superexponential. For its calculation we need
to know the dual pairing between partial derivatives and coordinates.
Explicitly, we have as non-vanishing expressions
\begin{equation}
\langle(\partial_{\theta})_{2},\theta^{2}\rangle_{{L},\bar{R}}=\langle
(\partial_{\theta})_{1},\theta^{1}\rangle_{{L},\bar{R}}=\langle(\partial
_{\theta})_{1}(\partial_{\theta})_{2},\theta^{2}\theta^{1}\rangle_{{L},\bar
{R}}=1,
\end{equation}
which follow from the very definition of the dual pairing together with the
action of partial derivatives on Grassmann variables. From the above result we
can at once read off theelements of the two bases being dual to each other.
Inserting these elements into the general formulae for the exponential in Eq.
(\ref{eform}) we obtain as explicit form of the q-deformed superexponential on
Manin plane
\begin{equation}
\mbox{exp}(\theta_{\bar{R}}\mid(\partial_{\theta})_{L})=1\otimes1+\theta
^{1}\otimes(\partial_{\theta})_{1}+\theta^{2}\otimes(\partial_{\theta}%
)_{2}+\theta^{2}\theta^{1}\otimes(\partial_{\theta})_{1}(\partial_{\theta
})_{2}.
\end{equation}
Repeating the same steps as before for the conjugated partial derivatives we
get instead
\begin{equation}
\langle(\hat{\partial}_{\theta})_{1},\theta^{1}\rangle_{\bar{L},{R}}%
=\langle(\hat{\partial}_{\theta})_{2},\theta^{2}\rangle_{\bar{L},{R}}%
=\langle(\hat{\partial}_{\theta})_{2}(\hat{\partial}_{\theta})_{1},\theta
^{1}\theta^{2}\rangle_{\bar{L},{R}}=1,
\end{equation}
and consequently
\begin{equation}
\mbox{exp}(\theta_{R}\mid(\hat{\partial}_{\theta})_{\bar{L}})=1\otimes
1+\theta^{1}\otimes(\hat{\partial}_{\theta})_{1}+\theta^{2}\otimes
(\hat{\partial}_{\theta})_{2}+\theta^{1}\theta^{2}\otimes(\hat{\partial
}_{\theta})_{2}(\hat{\partial}_{\theta})_{1}.
\end{equation}
The above two results together with those corresponding to unconjugated
partial derivatives establish the correspondences
\begin{align}
\langle\underline{\partial}_{\theta},\underline{\theta}\rangle_{{L},\bar{R}}
&  \overset{%
\genfrac{}{}{0pt}{}{\alpha\leftrightarrow\alpha^{\prime}}{q\leftrightarrow1/q}%
}{\longleftrightarrow}\langle\underline{\hat{\partial}}_{\theta}%
,\underline{\theta}\rangle_{\bar{L},R}\,,\label{eco2}\\
\mbox{exp}(\theta_{\bar{R}}\mid(\partial_{\theta})_{L})  &  \overset{%
\genfrac{}{}{0pt}{}{\alpha\leftrightarrow\alpha^{\prime}}{q\leftrightarrow1/q}%
}{\longleftrightarrow}\mbox{exp}(\theta_{{R}}\mid(\hat{\partial}_{\theta
})_{\bar{L}})\,,\nonumber
\end{align}
where the symbol $\overset{%
\genfrac{}{}{0pt}{}{\alpha\leftrightarrow\alpha^{\prime}}{q\leftrightarrow1/q}%
}{\longleftrightarrow}$ now denotes a transition via
\begin{equation}
\theta^{\alpha}\leftrightarrow\theta^{\alpha^{\prime}},\quad(\partial_{\theta
})_{\alpha}\leftrightarrow(\hat{\partial}_{\theta})_{\alpha^{\prime}},\quad
q\leftrightarrow q^{-1}.
\end{equation}

The above considerations on dual pairings and superexponentials are based on
the use of left derivatives, but they carry over to right derivatives as well
with a few necessary modifications. Towards this end, we have to realize that
the definitions
\begin{align}
\langle f(\underline{\theta}),g(\underline{\partial}_{\theta})\rangle
_{L,\bar{R}}  &  \equiv\varepsilon(f(\underline{\theta})\;\bar{\lhd
}\;g(\underline{\partial}_{\theta})),\\
\langle f(\underline{\theta}),g(\underline{\hat{\partial}}_{\theta}%
)\rangle_{\bar{L},{R}}  &  \equiv\varepsilon(f(\underline{\theta})\lhd
g(\underline{\hat{\partial}}_{\theta})),\nonumber
\end{align}
give a dual pairing as well. Now, we can repeat the same resonings as above.
This way we get%
\begin{align}
\langle\theta_{1},\partial_{\theta}^{1}\rangle_{L,\bar{R}}  &  =\langle
\theta_{2},\partial_{\theta}^{2}\rangle_{{L},\bar{R}}=\langle\theta_{1}%
\theta_{2},\partial_{\theta}^{2}\partial_{\theta}^{1}\rangle_{{L},\bar{R}%
}=1,\\
\langle\theta_{1},\hat{\partial}_{\theta}^{1}\rangle_{\bar{L},R}  &
=\langle\theta_{2},\hat{\partial}_{\theta}^{2}\rangle_{\bar{L},R}%
=\langle\theta_{2}\theta_{1},\hat{\partial}_{\theta}^{1}\hat{\partial}%
_{\theta}^{2}\rangle_{\bar{L},R}=1,\nonumber
\end{align}
and
\begin{align}
\mbox{exp}((\partial_{\theta})_{\bar{R}}\mid\theta_{L})  &  =1\otimes
1+\partial_{\theta}^{1}\otimes\theta_{1}+\partial_{\theta}^{2}\otimes
\theta_{2}+\partial^{2}\partial^{1}\otimes\theta_{1}\theta_{2},\\
\mbox{exp}((\hat{\partial}_{\theta})_{R}\mid\theta_{\bar{L}})  &
=1\otimes1+\hat{\partial}_{\theta}^{1}\otimes\theta_{1}+\hat{\partial}%
_{\theta}^{2}\otimes\theta_{2}+\hat{\partial}^{1}\hat{\partial}^{2}%
\otimes\theta_{2}\theta_{1}.\nonumber
\end{align}
Comparing these results to those for left derivatives shows us the existence
of the crossing symmetries
\begin{align}
&  \langle\underline{\theta},\underline{\partial}_{\theta}\rangle_{{L},\bar
{R}}\overset{\alpha\leftrightarrow\alpha^{\prime}}{\longleftrightarrow}%
\langle\underline{\partial}_{\theta},\underline{\theta}\rangle_{{L},\bar{R}%
}\;,\\
&  \langle\underline{\theta},\underline{\hat{\partial}}_{\theta}\rangle
_{\bar{L},R}\overset{\alpha\leftrightarrow\alpha^{\prime}}{\longleftrightarrow
}\langle\underline{\hat{\partial}}_{\theta},\underline{\theta}\rangle_{\bar
{L},R}\;,\nonumber
\end{align}
and
\begin{align}
\mbox{exp}((\partial_{\theta})_{\bar{R}}\mid\theta_{L})  &  \overset
{\alpha\leftrightarrow\alpha^{\prime}}{\longleftrightarrow}\mbox{exp}(\theta
_{\bar{R}}\mid(\partial_{\theta})_{L})\;,\\
\mbox{exp}((\hat{\partial}_{\theta})_{{R}}\mid\theta_{\bar{L}})  &
\overset{\alpha\leftrightarrow\alpha^{\prime}}{\longleftrightarrow
}\mbox{exp}(\theta_{{R}}\mid(\hat{\partial}_{\theta})_{\bar{L}})\;,\nonumber
\end{align}
where $\overset{\alpha\leftrightarrow\alpha^{\prime}}{\longleftrightarrow}$
indicates one of the following two substitutions:%
\begin{align}
\text{a) }(\partial_{\theta})_{\alpha}  &  \leftrightarrow\theta_{\alpha
},\quad\theta^{\alpha}\leftrightarrow\partial_{\theta}^{\alpha}\;,\\
\text{b) }(\hat{\partial}_{\theta})_{\alpha}  &  \leftrightarrow\theta
_{\alpha},\quad\theta^{\alpha}\leftrightarrow\hat{\partial}_{\theta}^{\alpha
}\;.\nonumber
\end{align}

Next, we would like to deal with Grassmann translations. As was pointed out in
Sec. \ref{Ideas} translations on quantum spaces are described by the
coproduct, which on spinor coordinates reads \cite{Mik04}
\begin{align}
\Delta_{\bar{L}}(\theta^{1})  &  =\theta^{1}\otimes1+\tilde{\Lambda}\tau
^{1/4}\otimes\theta^{1}+q^{-1}\lambda\tilde{\Lambda}\tau^{-1/4}T^{-}%
\otimes\theta^{2},\label{CopCon2dim}\\
\Delta_{\bar{L}}(\theta^{2})  &  =\theta^{2}\otimes1+\tilde{\Lambda}%
\tau^{-1/4}\otimes\theta^{2}.\nonumber
\end{align}
Notice that $\tau,$ $T^{+}$ and $\tilde{\Lambda}$ denote generators of
$U_{q}(su(2))$ and a scaling operator, respectively. Now, we can follow the
same reasonings already applied in Ref. \cite{Wac04-10}. In this manner the
coproduct is split into two parts by introducing left and right coordinates:%
\begin{equation}
\Delta_{\bar{L}}(\theta^{\alpha})=\theta^{\alpha}\otimes1+(\mathcal{\bar{L}%
}_{\theta})_{\beta}^{\alpha}\otimes\theta^{\beta}=\theta_{l}^{\alpha}%
+\theta_{r}^{\alpha}, \label{CoSinGen}%
\end{equation}
where
\begin{equation}
\theta_{l}^{\alpha}\equiv\theta_{l}^{\alpha}\otimes1,\quad\theta_{r}^{\alpha
}\equiv(\mathcal{\bar{L}}_{\theta})_{\beta}^{\alpha}\otimes\theta^{\beta
},\quad\alpha=1,2.
\end{equation}
Since the entries of the $\mathcal{L}$-matrix are built up out of symmetry
generators, the commutation relations between right and left coordinates can
be derived in a straightforward manner from the commutation relations between
symmetry generators and Grassmann variables (their explicit form was given in
Ref. \cite{Mik04}). It follows that
\begin{align}
\theta_{r}^{1}\theta_{l}^{1}  &  =-\theta_{l}^{1}\theta_{r}^{1}%
,\label{BraiRightLeft}\\
\theta_{r}^{1}\theta_{l}^{2}  &  =-q^{-1}\theta_{l}^{2}\theta_{r}^{1}%
-q^{-1}\lambda\theta_{l}^{1}\theta_{r}^{2},\nonumber\\
\theta_{r}^{2}\theta_{l}^{1}  &  =-q^{-1}\theta_{l}^{1}\theta_{r}%
^{2},\nonumber\\
\theta_{r}^{2}\theta_{l}^{2}  &  =-\theta_{l}^{2}\theta_{r}^{2}.\nonumber
\end{align}
Furthermore, these relations imply
\begin{align}
\Delta_{\bar{L}}(\theta^{2}\theta^{1})  &  =\Delta_{\bar{L}}(\theta
^{2})\,\Delta_{\bar{L}}(\theta^{1})=(\theta_{l}^{2}+\theta_{r}^{2})(\theta
_{l}^{1}+\theta_{r}^{1})\label{CoDoubGen}\\
&  =\theta_{l}^{2}\theta_{l}^{1}+\theta_{l}^{2}\theta_{r}^{1}+\theta_{r}%
^{2}\theta_{l}^{1}+\theta_{r}^{2}\theta_{r}^{1}\nonumber\\
&  =\theta_{l}^{2}\theta_{l}^{1}+\theta_{l}^{2}\theta_{r}^{1}-q^{-1}\theta
_{l}^{1}\theta_{r}^{2}+\theta_{r}^{2}\theta_{r}^{1}.\nonumber
\end{align}
Notice that in the last step we have switched all right coordinates to the
right of all left coordinates. Now, we are in a position to read off from the
above results the explicit form of a Grassmann translation. Specifically, it
becomes for a supernumber written in the form of Eq. (\ref{RepSup2dim})%
\begin{align}
f(\theta\underline{\oplus}_{\bar{L}}\psi)\equiv &  \left.  \,f(\theta_{l}%
^{2}+\theta_{r}^{2},\theta_{l}^{2}+\theta_{r}^{1})\right\vert _{\theta
_{l}^{\alpha}\rightarrow\,\theta^{\alpha},\,\theta_{r}^{\alpha}\rightarrow
\Psi^{\alpha}}\\
=  &  \big [f^{\prime}+f_{1}(\theta^{1}+\psi^{1})+f_{2}(\theta^{2}+\psi
^{2})\nonumber\\
&  +\;f_{21}(\theta_{l}^{2}+\theta_{r}^{2})(\theta_{l}^{2}+\theta_{r}%
^{1})\big ]_{\theta_{l}^{\alpha}\rightarrow\,\theta^{\alpha},\,\theta
_{r}^{\alpha}\rightarrow\Psi^{\alpha}}\nonumber\\
=  &  f^{\prime}+f_{1}(\theta^{1}+\psi^{1})+f_{2}(\theta^{2}+\psi
^{2})\nonumber\\
&  +\;f_{21}(\theta^{2}\theta^{1}+\theta^{2}\psi^{1}-q^{-1}\theta^{1}\psi
^{2}+\psi^{2}\psi^{1}).\nonumber
\end{align}
For a proper understanding of the definition above, one should realize that
substitutions can only be performed after all right coordinates have been
commuted to the right of an expression.

In order to introduce translations in the opposite direction we need to
consider the antipode corresponding to the coproducts in Eq. (\ref{CopCon2dim}%
). On spinor coordinates this antipode takes the form\ \cite{Mik04}
\begin{align}
S_{\bar{L}}(\theta^{1})  &  =-\tilde{\Lambda}^{-1}\tau^{-1/4}\theta^{1}%
+q^{-2}\lambda\tilde{\Lambda}^{-1}\tau^{-1/4}T^{-}\theta^{2},\\
S_{\bar{L}}(\theta^{2})  &  =\tilde{\Lambda}^{-1}\tau^{1/4}\theta
^{2}.\nonumber
\end{align}
However, what we are looking for is an antipode in terms of right or left
coordinates. To achieve this, we have to exploit the Hopf algebra axiom
\begin{equation}
m\circ(S\otimes id)\circ\Delta=m\circ(id\otimes S)\circ\Delta=\varepsilon,
\label{HopAx1}%
\end{equation}
where $m$ denotes multiplication in the Grassmann algebra. If we substitute
for the coproduct the expressions in Eqs. (\ref{CoSinGen}) and
(\ref{CoDoubGen}), we arrive at%
\begin{equation}
S_{\bar{L}}(\theta^{\alpha})+\theta^{\alpha}=\theta^{\alpha}+S_{\bar{L}%
}(\theta^{\alpha})=0,\quad\alpha=1,2,
\end{equation}
and%
\begin{align}
&  S_{\bar{L}}(\theta^{2}\theta^{1})+S_{\bar{L}}(\theta^{2})\,\theta
^{1}-q^{-1}S_{\bar{L}}(\theta^{1})\,\theta^{2}+\theta^{2}\theta^{1}\\
&  =\theta^{2}\theta^{1}+\theta^{2}S_{\bar{L}}(\theta^{1})-q^{-1}\theta
^{1}S_{\bar{L}}(\theta^{2})+S_{\bar{L}}(\theta^{2}\theta^{1})=0.\nonumber
\end{align}
This system of equations can be solved for $S_{\bar{L}}(\theta^{\alpha}),$
$\alpha=1,2,$ and $S_{\bar{L}}(\theta^{2}\theta^{1}),$ leaving us with%
\begin{align}
S_{\bar{L}}(\theta^{\alpha})  &  =-\theta^{\alpha},\quad\alpha=1,2,\\
S_{\bar{L}}(\theta^{2}\theta^{1})  &  =q^{-2}\theta^{2}\theta^{1}.\nonumber
\end{align}
Finally, the last results allow us to introduce the following operation on
supernumbers:
\begin{align}
f(\underline{\ominus}_{\bar{L}}\theta)  &  \equiv S_{\bar{L}}\,\left(
f(\theta^{2},\theta^{1})\right) \\
&  =f^{\prime}+f_{1}\,S_{\bar{L}}(\theta^{1})+f_{2}\,S_{\bar{L}}(\theta
^{2})+f_{21}\,S_{\bar{L}}(\theta^{2}\theta^{1})\nonumber\\
&  =f^{\prime}-f_{1}\theta^{1}-f_{2}\theta^{2}+q^{-2}f_{21}\theta^{2}%
\theta^{1}.\nonumber
\end{align}

Our considerations about translations can also be applied to the opposite Hopf
structure given by
\begin{equation}
\Delta_{\bar{R}}\equiv\tau\circ\Delta_{\bar{L}},\quad S_{\bar{R}}\equiv
S_{\bar{L}}^{-1},
\end{equation}
where $\tau$ denotes transposition of tensor factors. Right and left
coordinates are now defined by
\begin{equation}
\theta_{l}^{\alpha}\equiv\theta^{\beta}\otimes(\mathcal{\bar{L}}_{\theta
})_{\beta}^{\alpha},\quad\theta_{r}^{\alpha}\equiv1\otimes\theta^{\alpha}.
\end{equation}
Then we have for coproduct and antipode respectively%
\begin{align}
\Delta_{\bar{R}}(\theta^{\alpha})  &  =\theta_{l}^{\alpha}+\theta_{r}^{\alpha
},\quad\alpha=1,2,\\
\Delta_{\bar{R}}(\theta^{1}\theta^{2})  &  =\theta_{l}^{1}\theta_{l}%
^{2}+\theta_{l}^{1}\theta_{r}^{2}-q\theta_{l}^{2}\theta_{r}^{1}+\theta_{r}%
^{1}\theta_{r}^{2},\nonumber
\end{align}
and
\begin{align}
S_{\bar{R}}(\theta^{\alpha})  &  =-\theta^{\alpha},\quad\alpha=1,2,\\
S_{\bar{R}}(\theta^{1}\theta^{2})  &  =q^{2}\theta^{1}\theta^{2}.\nonumber
\end{align}
Consequently, we find as corresponding operations on supernumbers%
\begin{align}
f(\theta\underline{\oplus}_{\bar{R}}\psi)  &  =f^{\prime}+f_{1}(\theta
^{1}+\psi^{1})+f_{2}(\theta^{2}+\psi^{2})\\
&  +\;f_{12}(\theta^{1}\theta^{2}+\theta^{1}\psi^{2}-q\theta^{2}\psi^{1}%
+\psi^{1}\psi^{2}),\nonumber\\
f(\underline{\ominus}_{\bar{R}}\theta)  &  =f^{\prime}-f_{1}\theta^{1}%
-f_{2}\theta^{2}+q^{2}f_{12}\theta^{1}\theta^{2}.\nonumber
\end{align}
From what we have done so far we can easily derive the crossing symmetries
\begin{equation}
\Delta_{\bar{R}},S_{\bar{R}},\underline{\oplus}_{\bar{R}},\underline{\ominus
}_{\bar{R}}\overset{%
\genfrac{}{}{0pt}{}{\alpha\leftrightarrow\alpha^{\prime}}{q\leftrightarrow1/q}%
}{\longleftrightarrow}\Delta_{\bar{L}},S_{\bar{L}},\underline{\oplus}_{\bar
{L}},\underline{\ominus}_{\bar{L}},
\end{equation}
with the transition symbol having the same meaning as in Eq. (\ref{eco2}).

It would have been possible to begin with the other Hopf structure by starting
with the coproducts $\Delta_{L}$, $\Delta_{R}$ and the antipodes $S_{L}$,
$S_{R}.$ However, the expressions for the corresponding operations
$\underline{\oplus}_{{L}},$ $\underline{\ominus}_{L}$ and $\underline{\oplus
}_{R},$ $\underline{\ominus}_{R}$ are obtained most easily by the
transformation rules
\begin{align}
&  \Delta_{L},S_{L},\underline{\oplus}_{{L}},\underline{\ominus}_{{L}}%
\overset{%
\genfrac{}{}{0pt}{}{a\leftrightarrow\alpha^{\prime}}{q\leftrightarrow1/q}%
}{\longleftrightarrow}\Delta_{\bar{L}},S_{\bar{L}},\underline{\oplus}_{\bar
{L}},\underline{\ominus}_{\bar{L}},\\
&  \Delta_{R},S_{R},\underline{\oplus}_{R},\underline{\ominus}_{R}\overset{%
\genfrac{}{}{0pt}{}{\alpha\leftrightarrow\alpha^{\prime}}{q\leftrightarrow1/q}%
}{\longleftrightarrow}\Delta_{\bar{R}},S_{\bar{R}},\underline{\oplus}_{\bar
{R}},\underline{\ominus}_{\bar{R}},\nonumber
\end{align}
as can be proven by a direct calculation.

Our final comment concerns commutation relations between supernumbers and
arbitrary elements of other quantum spaces. In Sec. \ref{Ideas} we called
formulae for calculating such relations braided products. Recalling the
identities in (\ref{BraiBasTheta}), we can conclude that braided products for
supernumbers take on the form
\begin{align}
f(\theta^{2},\theta^{1})\;\underline{\odot}_{L}\,g  &  =g\otimes f^{\prime
}+f_{\alpha}\left(  (\mathcal{L}_{\theta})_{\beta}^{\alpha}\rhd g\right)
\otimes\theta^{\beta}\\
&  +\;f_{21}\left(  (\mathcal{L}_{\theta})_{\gamma}^{2}(\mathcal{L}_{\theta
})_{\delta}^{1}\rhd g\right)  \otimes\theta^{\gamma}\theta^{\delta
},\nonumber\\
g\;\underline{\odot}_{R}\,f(\theta^{2},\theta^{1})  &  =f^{\prime}\otimes
g+\theta^{\beta}\otimes\left(  g\lhd(\mathcal{L}_{\theta})_{\beta}^{\alpha
}\right)  \,f_{\alpha}\nonumber\\
&  +\;\left(  g\lhd(\mathcal{L}_{\theta})_{\gamma}^{2}(\mathcal{L}_{\theta
})_{\delta}^{1}\right)  \otimes\theta^{\gamma}\theta^{\delta},\nonumber
\end{align}
and likewise for the other braiding with the $\mathcal{L}$-operator now
substituted by its conjugate. Notice that in the last two identities summation
over all repeated indices is to be understood. After having inserted the
explicit form of the $\mathcal{L}$-operator and then rearranging, it follows
that (for compactness, we have abbreviated monomials of ordering $\theta
^{2}\theta^{1}$ by the symbol $\theta^{\underline{K}}$)%
\begin{align}
f(\theta^{2},\theta^{1})\;\underline{\odot}_{L/\bar{L}}\,g  &  =g\otimes
f^{\prime}+\sum\nolimits_{\underline{K}}\left(  (O_{f})_{L/\bar{L}%
}^{\underline{K}}\rhd g\right)  \otimes\theta^{\,\underline{K}}%
,\label{2dimBraiPro}\\
g\;\underline{\odot}_{R/\bar{R}}\,f(\theta^{2},\theta^{1})  &  =f^{\prime
}\otimes g+\sum\nolimits_{\underline{K}}\theta^{\,\underline{K}}\otimes\left(
g\lhd(O_{f})_{L/\bar{L}}^{\underline{K}}\right)  ,\nonumber
\end{align}
where
\begin{align}
(O_{f})_{L}^{1}  &  =\tilde{\Lambda}^{-1}\tau^{-1/4}(f_{1}-q\lambda f_{2}%
T^{+}),\\
(O_{f})_{L}^{2}  &  =\tilde{\Lambda}^{-1}\tau^{1/4}f_{2},\nonumber\\%
[0.1in]%
(O_{f})_{L}^{21}  &  =\tilde{\Lambda}^{-2},
\end{align}
and
\begin{align}
(O_{f})_{\hat{L}}^{1}  &  =\tilde{\Lambda}\tau^{1/4}f_{1},\\
(O_{f})_{\bar{L}}^{2}  &  =\tilde{\Lambda}\tau^{-1/4}(f_{2}+q^{-1}\lambda
f_{1}T^{-}),\nonumber\\%
[0.1in]%
(O_{f})_{\bar{L}}^{21}  &  =\tilde{\Lambda}^{2}.
\end{align}

\section{Three-Dimensional q-deformed Euclidean space\label{Sec3}}

The three-dimensional antisymmetrized Euclidean space (for its definition see
again Appendix \ref{AppQua}) can be treated in very much the same way as the
two-dimensional quantum plane. Thus we restrict ourselves to stating the
results, only. In complete analogy to the two-dimensional case we define left
and right Grassmann integrals by
\begin{align}
\int f(\theta^{+},\theta^{3},\theta^{-})\;d_{L}^{3}\theta &  \equiv
(\partial_{\theta})_{-}(\partial_{\theta})_{3}(\partial_{\theta})_{+}\rhd
f(\theta^{+},\theta^{3},\theta^{-})=f_{+3-},\\
\int d_{R}^{3}\theta\;f(\theta^{-},\theta^{3},\theta^{+})  &  \equiv
f(\theta^{-},\theta^{3},\theta^{+})\triangleleft(\hat{\partial}_{\theta}%
)_{+}(\hat{\partial}_{\theta})_{3}(\hat{\partial}_{\theta})_{-}=f_{-3+}%
,\nonumber
\end{align}
where the actions of partial derivatives have been calculated in Ref.
\cite{Mik04}. Again this definition has the consequence that
\begin{gather}
\int\theta^{+}\theta^{3}\theta^{-}\,d_{L}^{3}\theta=1,\quad\int1\;d_{L}%
^{3}\theta=\int\theta^{A}\,d_{L}^{3}\theta=0,\\
\int\theta^{+}\theta^{3}\,d_{L}^{3}\theta=\int\theta^{+}\theta^{-}\,d_{L}%
^{3}\theta=\int\theta^{3}\theta^{-}\,d_{L}^{3}\theta=0,\nonumber
\end{gather}
(we take the convention that labels which are not specified any further can
take on any of their possible values, i.e. in our case $A\in\{+,3,-\}$) and
\begin{gather}
\int d_{R}^{3}\theta\;\theta^{-}\theta^{3}\theta^{+}=1,\quad\int d_{R}%
^{3}\theta\;1\;=\int d_{R}^{3}\theta\;\theta^{A}=0,\\
\int d_{R}^{3}\theta\;\theta^{3}\theta^{+}=\int d_{R}^{3}\theta\;\,\theta
^{-}\theta^{+}=\int d_{R}^{3}\theta\;\theta^{-}\theta^{3}=0.\nonumber
\end{gather}
As usual, translational invariance is then given by
\begin{align}
\int(\partial_{\theta})_{A}\rhd f(\theta^{+},\theta^{3},\theta^{-})\;d_{L}%
^{3}\theta &  =0,\\
\int d_{R}^{3}\theta\;f(\theta^{-},\theta^{3},\theta^{+})\lhd(\hat{\partial
}_{\theta})_{A}  &  =0.\nonumber
\end{align}
The corresponding expressions arising from the conjugated differential
calculus are obtained from the above ones most easily by applying the
substitutions
\begin{align}
d_{L}^{3}\theta\leftrightarrow d_{\bar{L}}^{3}\theta,  &  \quad d_{R}%
^{3}\theta\leftrightarrow d_{\bar{R}}^{3}\theta,\\
f(\theta^{+},\theta^{3},\theta^{-})  &  \leftrightarrow f(\theta^{-}%
,\theta^{3},\theta^{+}),\nonumber\\
\theta^{A}\leftrightarrow\theta^{\bar{A}},  &  \quad(\partial_{\theta}%
)_{A}\leftrightarrow(\hat{\partial}_{\theta})_{\bar{A}},\nonumber\\
\rhd\leftrightarrow\bar{\rhd},  &  \quad\lhd\leftrightarrow\bar{\lhd
},\nonumber
\end{align}
where we have introduced indices with bar by setting
\begin{equation}
(\overline{+,3,-})=(-,3,+).
\end{equation}
For this substitution to become more clear we give as an example
\begin{equation}
\int(\partial_{\theta})_{+}\triangleright f(\theta^{+},\theta^{3},\theta
^{-})\,d_{L}^{3}\theta\leftrightarrow\int(\hat{\partial}_{\theta})_{-}%
\,\bar{\triangleright}\,f(\theta^{-},\theta^{3},\theta^{+})\,d_{\bar{L}}%
^{3}\theta.
\end{equation}

Next we come to the superexponentials. From the pairings
\begin{gather}
\langle(\partial_{\theta})_{-},\theta^{-}\rangle_{L,\bar{R}}=\langle
(\partial_{\theta})_{3},\theta^{3}\rangle_{L,\bar{R}}=\langle(\partial
_{\theta})_{+},\theta^{+}\rangle_{L,\bar{R}}=1,\\
\langle(\partial_{\theta})_{-}(\partial_{\theta})_{3},\theta^{3}\theta
^{-}\rangle_{L,\bar{R}}=1,\quad\langle(\partial_{\theta})_{-}(\partial
_{\theta})_{+},\theta^{+}\theta^{-}\rangle_{L,\bar{R}}=1,\nonumber\\
\langle(\partial_{\theta})_{3}(\partial_{\theta})_{+},\theta^{+}\theta
^{3}\rangle_{L,\bar{R}}=1,\quad\langle(\partial_{\theta})_{-}(\partial
_{\theta})_{3}(\partial_{\theta})_{+},\theta^{+}\theta^{3}\theta^{-}%
\rangle_{L,\bar{R}}=1,\nonumber
\end{gather}
we can read off for the exponential an expression which is the same as in the
undeformed case:
\begin{align}
&  \mbox{exp}(\theta_{\bar{R}}\mid(\partial_{\theta})_{L})\\
&  =1\otimes1+\theta^{+}\otimes(\partial_{\theta})_{+}+\theta^{3}%
\otimes(\partial_{\theta})_{3}\nonumber\\
&  +\,\theta^{-}\otimes(\partial_{\theta})_{-}+\theta^{+}\theta^{3}%
\otimes(\partial_{\theta})_{3}(\partial_{\theta})_{+}+\theta^{+}\theta
^{-}\otimes(\partial_{\theta})_{-}(\partial_{\theta})_{+}\nonumber\\
&  +\,\theta^{3}\theta^{-}\otimes(\partial_{\theta})_{-}(\partial_{\theta
})_{3}+\theta^{+}\theta^{3}\theta^{-}\otimes(\partial_{\theta})_{-}%
(\partial_{\theta})_{3}(\partial_{\theta})_{+}.\nonumber
\end{align}
In complete accordance with the considerations of the previous section we
found as crossing symmetries
\begin{align}
\langle\underline{\hat{\partial}}_{\theta},\underline{\theta}\rangle_{\bar
{L},R}  &  \overset{%
\genfrac{}{}{0pt}{}{+\leftrightarrow-}{q\leftrightarrow1/q}%
}{\longleftrightarrow}\langle\underline{{\partial}}_{\theta},{\theta}%
\rangle_{{L},\bar{R}}\;,\\
\langle\underline{\theta},\underline{\hat{\partial}}_{\theta}\rangle_{\bar
{L},R}  &  \overset{%
\genfrac{}{}{0pt}{}{+\leftrightarrow-}{q\leftrightarrow1/q}%
}{\longleftrightarrow}\langle\underline{\theta},\underline{\partial}_{\theta
}\rangle_{{L},\bar{R}}\;,\nonumber\\%
[0.1in]%
\langle\underline{\hat{\partial}}_{\theta},\underline{\theta}\rangle_{\bar
{L},R}  &  \overset{+\leftrightarrow-}{\longleftrightarrow}\langle
\underline{\theta},\underline{\hat{\partial}}_{\theta}\rangle_{\bar{L},R}\;,\\
\langle\underline{{\partial}}_{\theta},\underline{\theta}\rangle_{{L},\bar
{R}}  &  \overset{+\leftrightarrow-}{\longleftrightarrow}\langle
\underline{\theta},\underline{{\partial}}_{\theta}\rangle_{{L},\bar{R}%
}\;,\nonumber
\end{align}
and
\begin{align}
\mbox{exp}(\theta_{{R}}\mid(\hat{\partial}_{\theta})_{\bar{L}})  &  \overset{%
\genfrac{}{}{0pt}{}{+\leftrightarrow-}{q\leftrightarrow1/q}%
}{\longleftrightarrow}\mbox{exp}(\theta_{{\bar{R}}}\mid(\partial_{\theta}%
)_{L})\;,\\
\mbox{exp}((\hat{\partial}_{\theta})_{{R}}\mid\theta_{\bar{L}})  &  \overset{%
\genfrac{}{}{0pt}{}{+\leftrightarrow-}{q\leftrightarrow1/q}%
}{\longleftrightarrow}\mbox{exp}((\partial_{\theta})_{\bar{R}}\mid\theta
_{L})\;,\nonumber\\%
[0.1in]%
\mbox{exp}(\theta_{R}\mid(\hat{\partial}_{\theta})_{\bar{L}})  &
\overset{+\leftrightarrow-}{\longleftrightarrow}\mbox{exp}((\hat{\partial
}_{\theta})_{R}\mid\theta_{\bar{L}})\;,\\
\mbox{exp}(\theta_{\bar{R}}\mid(\partial_{\theta})_{{L}})  &  \overset
{+\leftrightarrow-}{\longleftrightarrow}\mbox{exp}((\partial_{\theta}%
)_{\bar{R}}\mid\theta_{L}).\nonumber
\end{align}
The symbol $\overset{%
\genfrac{}{}{0pt}{}{+\leftrightarrow-}{q\leftrightarrow1/q}%
}{\longleftrightarrow}$ now denotes a transition via
\begin{gather}
\theta^{A}\leftrightarrow\theta^{\bar{A}},\quad\theta_{A}\leftrightarrow
\theta_{\bar{A}},\quad q\leftrightarrow q^{-1},\\
(\partial_{\theta})^{A}\leftrightarrow(\hat{\partial}_{\theta})^{\bar{A}%
},\quad(\partial_{\theta})_{A}\leftrightarrow(\hat{\partial}_{\theta}%
)_{\bar{A}},\nonumber
\end{gather}
whereas $\overset{+\leftrightarrow-}{\longleftrightarrow}$ stands for one of
the following two substitutions:%
\begin{align}
\text{a) }\theta^{A}  &  \leftrightarrow\hat{\partial}^{A},\quad\hat{\partial
}_{A}\leftrightarrow\theta_{A},\\
\text{b) }\theta^{A}  &  \leftrightarrow{\partial}^{A},\quad{\partial}%
_{A}\leftrightarrow\theta_{A}.\nonumber
\end{align}

Now we concentrate our attention on the Hopf structure for Grassmann
variables. With its explicit form given in Ref.\ \cite{Mik04} we can show that
on a basis of normal ordered monomials the expressions for the coproduct
become
\begin{align}
\Delta_{L}(\theta^{A})  &  =\theta_{l}^{A}+\theta_{r}^{A},\\%
[0.1in]%
\Delta_{L}(\theta^{-}\theta^{3})  &  =\theta_{l}^{-}\theta_{l}^{3}+\theta
_{l}^{-}\theta_{l}^{3}-q^{2}\theta_{l}^{3}\theta_{r}^{-}+\theta_{r}^{-}%
\theta_{r}^{3},\\
\Delta_{L}(\theta^{-}\theta^{+})  &  =\theta_{l}^{-}\theta_{l}^{+}+\theta
_{l}^{-}\theta_{r}^{+}-q^{4}\theta_{l}^{+}\theta_{r}^{-}+\theta_{r}^{-}%
\theta_{r}^{+},\nonumber\\
\Delta_{L}(\theta^{3}\theta^{+})  &  =\theta_{l}^{3}\theta_{l}^{+}+\theta
_{l}^{3}\theta_{r}^{+}-q^{2}\theta_{l}^{+}\theta_{r}^{3}+\theta_{r}^{3}%
\theta_{r}^{+},\nonumber\\%
[0.1in]%
\Delta_{L}(\theta^{-}\theta^{3}\theta^{+})  &  =\theta_{l}^{-}\theta_{l}%
^{3}\theta_{l}^{+}+\theta_{l}^{-}\theta_{l}^{3}\theta_{r}^{+}+q^{6}\theta
_{l}^{3}\theta_{l}^{+}\theta_{r}^{-}-q^{2}\theta_{l}^{-}\theta_{l}^{+}%
\theta_{r}^{3}\\
&  +\;\theta_{l}^{-}\theta_{r}^{3}\theta_{r}^{+}-q^{2}\theta_{l}^{3}\theta
_{r}^{-}\theta_{r}^{+}+q^{6}\theta_{l}^{+}\theta_{r}^{-}\theta_{r}^{3}%
+\theta_{r}^{-}\theta_{r}^{3}\theta_{r}^{+}.\nonumber
\end{align}
Notice that right and left coordinates are defined in complete analogy to the
two-dimensional case. The above results are consistent with the antipodes
\begin{gather}
S_{L}(\theta^{A})=-\theta^{A},\\
S_{L}(\theta^{-}\theta^{3})=q^{4}\theta^{-}\theta^{3},\quad S_{L}(\theta
^{-}\theta^{+})=q^{4}\theta^{-}\theta^{+},\nonumber\\
S_{L}(\theta^{3}\theta^{+})=q^{4}\theta^{3}\theta^{+},\quad S_{L}(\theta
^{-}\theta^{3}\theta^{+})=q^{8}\theta^{-}\theta^{3}\theta^{+}.\nonumber
\end{gather}
Making use of the crossing symmetries
\begin{align}
&  \Delta_{L},S_{L}\overset{%
\genfrac{}{}{0pt}{}{+\leftrightarrow-}{q\leftrightarrow1/q}%
}{\longleftrightarrow}\Delta_{\bar{L}},S_{\bar{L}},\;\\
&  \Delta_{R},S_{R}\overset{%
\genfrac{}{}{0pt}{}{+\leftrightarrow-}{q\leftrightarrow1/q}%
}{\longleftrightarrow}\Delta_{\bar{R}},S_{\bar{R}}\;,\nonumber\\%
[0.1in]%
&  \Delta_{L},S_{L}\overset{%
\genfrac{}{}{0pt}{}{+\leftrightarrow-}{q\leftrightarrow1/q}%
}{\longleftrightarrow}\Delta_{R},S_{R}\;,\\
&  \Delta_{\bar{L}},S_{\bar{L}}\overset{%
\genfrac{}{}{0pt}{}{+\leftrightarrow-}{q\leftrightarrow1/q}%
}{\longleftrightarrow}\Delta_{\bar{R}},S_{\bar{R}}\;,\nonumber
\end{align}
we are able to find the corresponding expressions for the other types of Hopf
structures. Now, the explicit form of the q-deformed\ addition law for
supernumbers should be rather apparent from what we have done so far. Thus it
is left to the reader to write down the explicit form of the operations
$\underline{\oplus}$ and $\underline{\ominus}$.

We conclude this section by presenting explicit formulae for braided products
concerning supernumbers represented in the form of Eq. (\ref{SupNumAllg}),
where $\theta^{\,\underline{K}}$ shall now denote monomials of ordering
$\theta^{+}\theta^{3}\theta^{-}$. Explicitly, we have
\begin{align}
f(\theta^{+},\theta^{3},\theta^{-})\,\underline{\odot}_{L/\bar{L}}\,g  &
=g\otimes f^{\prime}+\sum\nolimits_{\underline{K}}\left(  (O_{f})_{L/\bar{L}%
}^{\underline{K}}\rhd g\right)  \otimes\theta^{\,\underline{K}},\\
g\,\underline{\odot}_{R/\bar{R}}\,f(\theta^{+},\theta^{3},\theta^{-})  &
=f^{\prime}\otimes g+\sum\nolimits_{\underline{K}}\theta^{\,\underline{K}%
}\otimes\left(  g\lhd(O_{f})_{L/\bar{L}}^{\underline{K}}\right)  ,\nonumber
\end{align}
where we introduced abbreviations for the following combinations of
$U_{q}(su_{2})$-generators:%
\begin{align}
(O_{f})_{L}^{+}  &  =\tilde{\Lambda}^{-1}f_{+}\tau^{1/2},\\
(O_{f})_{L}^{3}  &  =\tilde{\Lambda}^{-1}(f_{3}+q\lambda\lambda_{+}f_{+}%
\tau^{1/2}L^{+}),\nonumber\\
(O_{f})_{L}^{-}  &  =\tilde{\Lambda}^{-1}(f_{-}\tau^{1/2}+\lambda\lambda
_{+}f_{3}L^{+}+q^{2}\lambda^{2}\lambda_{+}f_{+}\tau^{1/2}(L^{+})^{2}%
),\nonumber\\%
[0.1in]%
(O_{f})_{L}^{+3}  &  =\tilde{\Lambda}^{-2}f_{+3}\tau^{1/2},\\
(O_{f})_{L}^{+-}  &  =\tilde{\Lambda}^{-2}(f_{+-}+q^{2}\lambda\lambda
_{+}f_{+3}\tau^{1/2}L^{+}),\nonumber\\
(O_{f})_{L}^{3-}  &  =\tilde{\Lambda}^{-2}(f_{3-}\tau^{-1/2}-q^{-1}%
\lambda\lambda_{+}f_{+-}L^{+}\nonumber\\
&  +\;q^{2}\lambda^{2}\lambda_{+}f_{+3}\tau^{1/2}(L^{+})^{2}),\nonumber\\%
[0.1in]%
(O_{f})_{L}^{+3-}  &  =\tilde{\Lambda}^{-3}f_{+3-},
\end{align}
and likewise for the second braiding
\begin{align}
(O_{f})_{\bar{L}}^{+}  &  =\tilde{\Lambda}(f_{+}\tau^{-1/2}+\lambda\lambda
_{+}f_{3}L^{-}+q^{-2}\lambda^{2}\lambda_{+}f_{-}\tau^{1/2}(L^{-})^{2}),\\
(O_{f})_{\bar{L}}^{3}  &  =\tilde{\Lambda}(f_{3}+q^{-1}\lambda\lambda_{+}%
f_{-}\tau^{1/2}L^{-}),\nonumber\\
(O_{f})_{\bar{L}}^{-}  &  =\tilde{\Lambda}f_{-}\tau^{-1/2},\nonumber\\%
[0.1in]%
(O_{f})_{\bar{L}}^{+3}  &  =\tilde{\Lambda}^{2}(f_{+3}\tau^{-1/2}%
-q^{-1}\lambda\lambda_{+}f_{+-}L^{-}\\
&  +\;q^{-2}\lambda^{2}\lambda_{+}f_{3-}\tau^{1/2}(L^{-})^{2}),\nonumber\\
(O_{f})_{\bar{L}}^{+-}  &  =\tilde{\Lambda}^{2}(f_{+-}+\lambda\lambda
_{+}f_{3-}\tau^{1/2}L^{-}),\nonumber\\
(O_{f})_{\bar{L}}^{3-}  &  =\tilde{\Lambda}^{2}f_{3-}\tau^{1/2},\nonumber\\%
[0.1in]%
(O_{f})_{\bar{L}}^{+3-}  &  =\tilde{\Lambda}^{3}f_{+3-}.
\end{align}
As in the two-dimensional case our formulae for braided products require to
know the actions of symmetry generators on quantum space elements. The
explicit form of these acions is already known from Refs. \cite{BW01} and
\cite{Mik04}.

\section{Four-Dimensional q-deformed Euclidean space\label{Sec4}}

All considerations of the previous sections pertain equally to the
antisymmetrized Euclidean space with four dimensions. For its definition and
some basic results used in the following we refer the reader to Appendix
\ref{AppQua} and Ref.\ \cite{Mik04}. To begin, we introduce Grassmann
integrals by
\begin{align}
&  \int f(\theta^{4},\theta^{3},\theta^{2},\theta^{1})\;d_{L}^{4}\theta\\
&  \equiv(\partial_{\theta})_{1}(\partial_{\theta})_{2}(\partial_{\theta}%
)_{3}(\partial_{\theta})_{4}\rhd f(\theta^{4},\theta^{3},\theta^{2},\theta
^{1})=f_{1234},\nonumber\\%
[0.1in]%
&  \int d_{R}^{4}\theta\,f(\theta^{1},\theta^{2},\theta^{3},\theta^{4})\\
&  \equiv f(\theta^{1},\theta^{2},\theta^{3},\theta^{4})\lhd(\hat{\partial
}_{\theta})_{4}(\hat{\partial}_{\theta})_{3}(\hat{\partial}_{\theta})_{3}%
(\hat{\partial}_{\theta})_{2}(\hat{\partial}_{\theta})_{1}=f_{4321}.\nonumber
\end{align}
Using the explicit form for the action of partial derivatives (see
Ref.\ \cite{Mik04}), one immediately arrives at
\begin{align}
&  \int\theta^{4}\theta^{3}\theta^{2}\theta^{1}\,d_{L}^{4}\theta=1,\quad
\int\theta^{i}\,d_{L}^{4}\theta=0,\quad i=1,\ldots,4,\\%
[0.1in]%
&  \int\theta^{i}\theta^{j}\,d_{L}^{4}\theta=0,\quad(i,j)\in
\{(4,3),(4,2),(4,1),(3,2),(3,1),(2,1)\},\nonumber\\%
[0.1in]%
&  \int\theta^{k}\theta^{l}\theta^{m}\,d_{L}^{4}\theta=0,\quad(k,l,m)\in
\{(4,3,2),(4,3,1),(4,2,1),(3,2,1)\},\nonumber
\end{align}
and
\begin{align}
&  \int d_{R}^{4}\theta\;\theta^{1}\theta^{2}\theta^{3}\theta^{4}=1,\quad\int
d_{R}^{4}\theta\;\theta^{i}=0,\\%
[0.1in]%
&  \int d_{R}^{4}\theta\;\theta^{i}\theta^{j}=0,\quad(i,j)\in
\{(1,2),(1,3),(1,4),(2,3),(2,4),(3,4)\},\nonumber\\%
[0.1in]%
&  \int d_{R}^{4}\theta\;\theta^{k}\theta^{l}\theta^{m}=0,\quad(k,l,m)\in
\{(1,2,3),(1,2,4),(1,3,4),(2,3,4)\}.\nonumber
\end{align}
In the same way one readily proves translational invariance, i.e.
\begin{align}
\int(\partial_{\theta})_{i}\rhd f(\theta^{4},\theta^{3},\theta^{2},\theta
^{1})\,d_{L}^{4}\theta &  =0,\\
\int d_{R}^{4}\theta\,f(\theta^{1},\theta^{2},\theta^{3},\theta^{4})\lhd
(\hat{\partial}_{\theta})_{i}  &  =0.\nonumber
\end{align}
Applying the substitutions
\begin{gather}
d_{L}^{4}\theta\leftrightarrow d_{\bar{L}}^{4}\theta,\quad d_{R}^{4}%
\theta\leftrightarrow d_{\bar{R}}^{4}\theta,\\
f(\theta^{1},\theta^{2},\theta^{3},\theta^{4})\leftrightarrow f(\theta
^{4},\theta^{3},\theta^{2},\theta^{1}),\nonumber\\
\theta^{i}\leftrightarrow\theta^{i^{\prime}},\quad(\partial_{\theta}%
)_{i}\leftrightarrow(\hat{\partial}_{\theta})_{i^{\prime}},\quad i^{\prime
}\equiv i-5,\nonumber\\
\rhd\leftrightarrow\bar{\rhd},\quad\lhd\leftrightarrow\bar{\lhd},\nonumber
\end{gather}
to all of the above expressions yields the corresponding identities for the
conjugated differential calculus.

Next, we turn to superexponentials. From the dual pairings
\begin{align}
&  \langle(\partial_{\theta})_{i},\theta^{i}\rangle_{{L},\bar{R}}=1,\quad
i=1\ldots,4,\\%
[0.1in]%
&  \langle(\partial_{\theta})_{1}(\partial_{\theta})_{2},\theta^{2}\theta
^{1}\rangle_{{L},\bar{R}}=\langle(\partial_{\theta})_{1}(\partial_{\theta
})_{3},\theta^{3}\theta^{1}\rangle_{{L},\bar{R}}\\
&  =\langle(\partial_{\theta})_{1}(\partial_{\theta})_{4},\theta^{4}\theta
^{1}\rangle_{{L},\bar{R}}=\langle(\partial_{\theta})_{2}(\partial_{\theta
})_{3},\theta^{3}\theta^{2}\rangle_{{L},\bar{R}}\nonumber\\
&  =\langle(\partial_{\theta})_{2}(\partial_{\theta})_{4},\theta^{4}\theta
^{2}\rangle_{{L},\bar{R}}=\langle(\partial_{\theta})_{3}(\partial_{\theta
})_{4},\theta^{4}\theta^{3}\rangle_{{L},\bar{R}}=1,\nonumber\\%
[0.1in]%
&  \langle(\partial_{\theta})_{1}(\partial_{\theta})_{2}(\partial_{\theta
})_{3},\theta^{3}\theta^{2}\theta^{1}\rangle_{{L},\bar{R}}=\langle
(\partial_{\theta})_{1}(\partial_{\theta})_{2}(\partial_{\theta})_{4}%
,\theta^{4}\theta^{2}\theta^{1}\rangle_{{L},\bar{R}}\\
&  =\langle(\partial_{\theta})_{1}(\partial_{\theta})_{3}(\partial_{\theta
})_{4},\theta^{4}\theta^{3}\theta^{1}\rangle_{{L},\bar{R}}=\langle
(\partial_{\theta})_{2}(\partial_{\theta})_{3}(\partial_{\theta})_{4}%
,\theta^{4}\theta^{3}\theta^{2}\rangle_{{L},\bar{R}}=1,\nonumber\\%
[0.1in]%
&  \langle(\partial_{\theta})_{1}(\partial_{\theta})_{2}(\partial_{\theta
})_{3}(\partial_{\theta})_{4},\theta^{4}\theta^{3}\theta^{2}\theta^{1}%
\rangle_{{L},\bar{R}}=1,
\end{align}
we can deduce for the exponential
\begin{align}
&  \mbox{exp}(\theta_{\bar{R}}\mid(\partial_{\theta})_{L})\\
&  =1\otimes1+\theta^{1}\otimes(\partial_{\theta})_{1}+\theta^{2}%
\otimes(\partial_{\theta})_{2}\nonumber\\
&  +\;\theta^{3}\otimes(\partial_{\theta})_{3}+\theta^{4}\otimes
(\partial_{\theta})_{4}+\theta^{4}\theta^{3}\otimes(\partial_{\theta}%
)_{3}(\partial_{\theta})_{4}\nonumber\\
&  +\;\theta^{4}\theta^{2}\otimes(\partial_{\theta})_{2}(\partial_{\theta
})_{4}+\theta^{4}\theta^{1}\otimes(\partial_{\theta})_{1}(\partial_{\theta
})_{4}\nonumber\\
&  +\;\theta^{3}\theta^{1}\otimes(\partial_{\theta})_{1}(\partial_{\theta
})_{3}+\theta^{2}\theta^{1}\otimes(\partial_{\theta})_{1}(\partial_{\theta
})_{2}\nonumber\\
&  +\;\theta^{4}\theta^{3}\theta^{2}\otimes(\partial_{\theta})_{2}%
(\partial_{\theta})_{3}(\partial_{\theta})_{4}+\theta^{4}\theta^{3}\theta
^{1}\otimes(\partial_{\theta})_{1}(\partial_{\theta})_{3}(\partial_{\theta
})_{4}\nonumber\\
&  +\;\theta^{4}\theta^{2}\theta^{1}\otimes(\partial_{\theta})_{1}%
(\partial_{\theta})_{2}(\partial_{\theta})_{4}+\theta^{3}\theta^{2}\theta
^{1}\otimes(\partial_{\theta})_{1}(\partial_{\theta})_{2}(\partial_{\theta
})_{3}\nonumber\\
&  +\;\theta^{4}\theta^{3}\theta^{2}\theta^{1}\otimes(\partial_{\theta}%
)_{1}(\partial_{\theta})_{2}(\partial_{\theta})_{3}(\partial_{\theta}%
)_{4}.\nonumber
\end{align}
The other types of pairings and exponentials correspond to the above
expressions through
\begin{align}
\langle\underline{\hat{\partial}}_{\theta},\underline{\theta}\rangle_{\bar
{L},R}  &  \overset{%
\genfrac{}{}{0pt}{}{i\leftrightarrow i^{\prime}}{q\leftrightarrow1/q}%
}{\longleftrightarrow}\langle\underline{{\partial}}_{\theta},{\theta}%
\rangle_{{L},\bar{R}}\;,\\
\langle\underline{\theta},\underline{\hat{\partial}}_{\theta}\rangle_{\bar
{L},R}  &  \overset{%
\genfrac{}{}{0pt}{}{i\leftrightarrow i^{\prime}}{q\leftrightarrow1/q}%
}{\longleftrightarrow}\langle\underline{\theta},\underline{\partial}_{\theta
}\rangle_{{L},\bar{R}}\;,\nonumber\\%
[0.1in]%
\langle\underline{\hat{\partial}}_{\theta},\underline{\theta}\rangle_{\bar
{L},R}  &  \overset{i\leftrightarrow i^{\prime}}{\longleftrightarrow}%
\langle\underline{\theta},\underline{\hat{\partial}}_{\theta}\rangle_{\bar
{L},R}\;,\\
\langle\underline{{\partial}}_{\theta},\underline{\theta}\rangle_{{L},\bar
{R}}  &  \overset{i\leftrightarrow i^{\prime}}{\longleftrightarrow}%
\langle\underline{\theta},\underline{{\partial}}_{\theta}\rangle_{{L},\bar{R}%
}\;,\nonumber
\end{align}
and
\begin{align}
\mbox{exp}(\theta_{{R}}\mid(\hat{\partial}_{\theta})_{\bar{L}})  &  \overset{%
\genfrac{}{}{0pt}{}{i\leftrightarrow i^{\prime}}{q\leftrightarrow1/q}%
}{\longleftrightarrow}\mbox{exp}(\theta_{{\bar{R}}}\mid(\partial_{\theta}%
)_{L})\;,\\
\mbox{exp}((\hat{\partial}_{\theta})_{{R}}\mid\theta_{\bar{L}})  &  \overset{%
\genfrac{}{}{0pt}{}{i\leftrightarrow i^{\prime}}{q\leftrightarrow1/q}%
}{\longleftrightarrow}\mbox{exp}((\partial_{\theta})_{\bar{R}}\mid\theta
_{L})\;,\nonumber\\%
[0.1in]%
\mbox{exp}(\theta_{R}\mid(\hat{\partial}_{\theta})_{\bar{L}})  &
\overset{i\leftrightarrow i^{\prime}}{\longleftrightarrow}\mbox{exp}((\hat
{\partial}_{\theta})_{R}\mid\theta_{\bar{L}})\;,\\
\mbox{exp}(\theta_{\bar{R}}\mid(\partial_{\theta})_{{L}})  &  \overset
{i\leftrightarrow i^{\prime}}{\longleftrightarrow}\mbox{exp}((\partial
_{\theta})_{\bar{R}}\mid\theta_{L}),\nonumber
\end{align}
where $\overset{%
\genfrac{}{}{0pt}{}{+\leftrightarrow-}{q\leftrightarrow1/q}%
}{\longleftrightarrow}$ symbolizes the transition
\begin{gather}
\theta^{i}\leftrightarrow\theta^{i^{\prime}},\quad\theta_{i}\leftrightarrow
\theta_{i^{\prime}},\quad q\leftrightarrow q^{-1},\\
(\partial_{\theta})^{i}\leftrightarrow(\hat{\partial}_{\theta})^{i^{\prime}%
},\quad(\partial_{\theta})_{i}\leftrightarrow(\hat{\partial}_{\theta
})_{i^{\prime}},\nonumber
\end{gather}
and $\overset{+\leftrightarrow-}{\longleftrightarrow}$ stands for one of the
following two substitutions:
\begin{align}
\text{a)\thinspace}\theta^{i}  &  \leftrightarrow\hat{\partial}^{i},\quad
\hat{\partial}_{i}\leftrightarrow\theta_{i},\\
\text{b)\thinspace}\theta^{i}  &  \leftrightarrow{\partial}^{i},\quad
{\partial}_{i}\leftrightarrow\theta_{i}.\nonumber
\end{align}

As we already know, Grassmann translations are determined by the explicit form
of the coproduct, for which we found in terms of right and left coordinates%
\begin{align}
\Delta_{L}(\theta^{i})  &  =\theta_{l}^{i}+\theta_{r}^{j},\quad i=1,...,4,\\%
[0.1in]%
\Delta_{L}(\theta^{j}\theta^{k})  &  =\theta_{l}^{j}\theta_{l}^{k}+\theta
_{l}^{j}\theta_{r}^{k}-q\theta_{l}^{k}\theta_{r}^{j}+\theta_{r}^{j}\theta
_{r}^{k}\;,\nonumber\\
\Delta_{L}(\theta^{1}\theta^{4})  &  =\theta_{l}^{1}\theta_{l}^{4}+\theta
_{l}^{1}\theta_{r}^{4}-q^{2}\theta_{l}^{4}\theta_{r}^{1}+\theta_{r}^{1}%
\theta_{r}^{4}\;,\nonumber\\
\Delta_{L}(\theta^{2}\theta^{3})  &  =\theta_{l}^{2}\theta_{l}^{3}+\theta
_{l}^{2}\theta_{r}^{3}-q^{2}\theta_{l}^{3}\theta_{r}^{2}+\theta_{r}^{2}%
\theta_{r}^{3}-q^{2}\lambda\theta_{l}^{4}\theta_{r}^{1}\;,\nonumber\\%
[0.1in]%
\Delta_{L}(\theta^{1}\theta^{2}\theta^{3})  &  =\theta_{l}^{1}\theta_{l}%
^{2}\theta_{l}^{3}+\theta_{l}^{1}\theta_{l}^{2}\theta_{r}^{3}-q^{2}\theta
_{l}^{1}\theta_{l}^{3}\theta_{r}^{2}\\
&  +\;q^{2}\theta_{l}^{2}\theta_{l}^{3}\theta_{r}^{1}+\theta_{l}^{1}\theta
_{r}^{2}\theta_{r}^{3}-q\theta_{l}^{2}\theta_{r}^{1}\theta_{r}^{3}\nonumber\\
&  +\;q^{3}\theta_{l}^{3}\theta_{r}^{1}\theta_{r}^{2}+\theta_{r}^{1}\theta
_{r}^{2}\theta_{r}^{3}-q^{2}\lambda\theta_{l}^{1}\theta_{l}^{4}\theta_{r}%
^{1}\;,\nonumber\\
\Delta_{L}(\theta^{1}\theta^{a}\theta^{4})  &  =\theta_{l}^{1}\theta_{l}%
^{a}\theta_{l}^{4}+\theta_{l}^{1}\theta_{l}^{a}\theta_{r}^{4}-q\theta_{l}%
^{1}\theta_{l}^{4}\theta_{r}^{a}\nonumber\\
&  +\;q^{3}\theta_{l}^{a}\theta_{l}^{4}\theta_{r}^{1}-q\theta_{l}^{a}%
\theta_{r}^{1}\theta_{r}^{4}+q^{3}\theta_{l}^{4}\theta_{r}^{1}\theta_{r}%
^{a}\nonumber\\
&  +\;\theta_{l}^{1}\theta_{r}^{a}\theta_{r}^{4}+\theta_{r}^{1}\theta_{r}%
^{a}\theta_{r}^{4}\;,\nonumber\\
\Delta_{L}(\theta^{2}\theta^{3}\theta^{4})  &  =\theta_{l}^{2}\theta_{l}%
^{3}\theta_{l}^{4}+\theta_{l}^{2}\theta_{l}^{3}\theta_{r}^{4}-q^{2}\theta
_{l}^{3}\theta_{r}^{2}\theta_{r}^{4}\nonumber\\
&  +\;q^{2}\theta_{l}^{4}\theta_{r}^{2}\theta_{r}^{3}+\theta_{l}^{2}\theta
_{r}^{3}\theta_{r}^{4}-q\theta_{l}^{2}\theta_{l}^{4}\theta_{r}^{3}\nonumber\\
&  +\;q^{3}\theta_{l}^{3}\theta_{l}^{4}\theta_{r}^{2}+\theta_{r}^{2}\theta
_{r}^{3}\theta_{r}^{4}-q^{2}\lambda\theta_{l}^{4}\theta_{r}^{1}\theta_{r}%
^{4}\;,\nonumber\\%
[0.1in]%
\Delta_{L}(\theta^{1}\theta^{2}\theta^{3}\theta^{4})  &  =\theta_{l}^{1}%
\theta_{l}^{2}\theta_{l}^{3}\theta_{l}^{4}+\theta_{l}^{1}\theta_{l}^{2}%
\theta_{l}^{3}\theta_{r}^{4}-q\theta_{l}^{1}\theta_{l}^{2}\theta_{l}^{4}%
\theta_{r}^{3}\\
&  +\;q^{3}\theta_{l}^{1}\theta_{l}^{3}\theta_{l}^{4}\theta_{r}^{2}%
-q^{4}\theta_{l}^{2}\theta_{l}^{3}\theta_{l}^{4}\theta_{r}^{1}+\theta_{l}%
^{1}\theta_{l}^{2}\theta_{r}^{3}\theta_{r}^{4}\nonumber\\
&  -\;q^{2}\theta_{l}^{1}\theta_{l}^{3}\theta_{r}^{2}\theta_{r}^{4}%
+q^{2}\theta_{l}^{1}\theta_{l}^{4}\theta_{r}^{2}\theta_{r}^{3}+q^{2}\theta
_{l}^{2}\theta_{l}^{3}\theta_{r}^{1}\theta_{r}^{4}\nonumber\\
&  -\;q^{4}\theta_{l}^{2}\theta_{l}^{4}\theta_{r}^{1}\theta_{r}^{3}%
+q^{6}\theta_{l}^{3}\theta_{l}^{4}\theta_{r}^{1}\theta_{r}^{2}+\theta_{l}%
^{1}\theta_{r}^{2}\theta_{r}^{3}\theta_{r}^{4}\nonumber\\
&  -\;q\theta_{l}^{2}\theta_{r}^{1}\theta_{r}^{3}\theta_{r}^{4}+q^{3}%
\theta_{l}^{3}\theta_{r}^{1}\theta_{r}^{2}\theta_{r}^{4}-q^{4}\theta_{l}%
^{4}\theta_{r}^{1}\theta_{r}^{2}\theta_{r}^{3}\nonumber\\
&  +\;\theta_{r}^{1}\theta_{r}^{2}\theta_{r}^{3}\theta_{r}^{4}-q^{2}%
\lambda\theta_{l}^{1}\theta_{l}^{4}\theta_{r}^{1}\theta_{r}^{4},\nonumber
\end{align}
where
\[
a=2,3\,,\quad(j,k)\in{\{}(1,2),(1,3),(2,4),(3,4){\}}.
\]
For the sake of completeness we wish to write down the expressions the
accompanying antipode gives on the same basis of normal ordered monomials:
\begin{align}
&  S_{L}(\theta^{i})=-\theta^{i},\quad i=1,\ldots,4,\\
&  S_{L}(\theta^{j}\theta^{k})=q^{2}\theta^{j}\theta^{k},\nonumber\\
&  S_{L}(\theta^{l}\theta^{m}\theta^{n})=-q^{6}\theta^{l}\theta^{m}\theta
^{n},\nonumber\\
&  S_{L}(\theta^{1}\theta^{2}\theta^{3}\theta^{4})=q^{12}\theta^{1}\theta
^{2}\theta^{3}\theta^{4},\nonumber
\end{align}
with%
\begin{align}
(i,j)  &  \in\{(1,2),(1,3),(1,4),(2,3),(2,4),(3,4)\},\\
(l,m,n)  &  \in\{(1,2,3),(1,2,4),(1,3,4),(2,3,4)\}.\nonumber
\end{align}
The formulae for the other types of Hopf structures follow from
\begin{align}
&  \Delta_{L},S_{L}\overset{%
\genfrac{}{}{0pt}{}{i\leftrightarrow i^{\prime}}{q\leftrightarrow1/q}%
}{\longleftrightarrow}\Delta_{\bar{L}},S_{\bar{L}},\\
&  \Delta_{R},S_{R}\overset{%
\genfrac{}{}{0pt}{}{i\leftrightarrow i^{\prime}}{q\leftrightarrow1/q}%
}{\longleftrightarrow}\Delta_{\bar{R}},S_{\bar{R}},\nonumber\\%
[0.1in]%
&  \Delta_{L},S_{L}\overset{%
\genfrac{}{}{0pt}{}{i\leftrightarrow i^{\prime}}{q\leftrightarrow1/q}%
}{\longleftrightarrow}\Delta_{R},S_{R},\\
&  \Delta_{\bar{L}},S_{\bar{L}}\overset{%
\genfrac{}{}{0pt}{}{i\leftrightarrow i^{\prime}}{q\leftrightarrow1/q}%
}{\longleftrightarrow}\Delta_{\bar{R}},S_{\bar{R}}.\nonumber
\end{align}

Last but not least we list expressions for braided products with supernumbers.
In general, we have
\begin{align}
f(\theta^{1},\theta^{2},\theta^{3},\theta^{4})\,\underline{\odot}_{L/\bar{L}%
}\,g  &  =g\otimes f^{\prime}+\sum\nolimits_{\underline{K}}\left(
(O_{f})_{L/\bar{L}}^{\underline{K}}\rhd g\right)  \otimes\theta^{\,\underline
{K}},\\
g\,\underline{\odot}_{R/\bar{R}}\,f(\theta^{1},\theta^{2},\theta^{3}%
,\theta^{4})  &  =f^{\prime}\otimes g+\sum\nolimits_{\underline{K}}%
\theta^{\,\underline{K}}\otimes\left(  g\lhd(O_{f})_{L/\bar{L}}^{\underline
{K}}\right)  ,\nonumber
\end{align}
where the sum includes all monomials of ordering $\theta^{1}\theta^{2}%
\theta^{3}\theta^{4}$, and the operators we introduced in the above formulae
are specified by the following combinations of $U_{q}(so_{4})$-generators (for
their action on quantum space elements we refer again to Refs.\ \cite{BW01}
and \cite{Mik04}):%
\begin{align}
(O_{f})_{L}^{1}  &  =\tilde{\Lambda}K_{1}^{1/2}K_{2}^{1/2}(f_{1}+q\lambda
f_{2}L_{1}^{+}+q\lambda f_{3}L_{2}^{+}\\
&  -\;q^{2}\lambda^{2}f_{4}L_{1}^{+}L_{2}^{+}),\nonumber\\
(O_{f})_{L}^{2}  &  =\tilde{\Lambda}K_{1}^{-1/2}K_{2}^{1/2}(f_{2}-q\lambda
f_{4}L_{2}^{+}),\nonumber\\
(O_{f})_{L}^{3}  &  =\tilde{\Lambda}K_{1}^{1/2}K_{2}^{-1/2}(f_{3}-q\lambda
f_{4}L_{1}^{+}),\nonumber\\
(O_{f})_{L}^{4}  &  =\tilde{\Lambda}K_{1}^{-1/2}K_{2}^{-1/2}f_{4},\nonumber\\%
[0.1in]%
(O_{f})_{L}^{12}  &  =\tilde{\Lambda}^{2}K_{2}(f_{12}-q\lambda f_{14}L_{2}%
^{+}-q^{2}\lambda f_{23}L_{2}^{+}\\
&  -\;q\lambda^{2}f_{34}(L_{2}^{+})^{2}),\nonumber\\
(O_{f})_{L}^{13}  &  =\tilde{\Lambda}^{2}K_{1}(f_{13}-q\lambda f_{14}L_{1}%
^{+}+\lambda f_{23}L_{1}^{+}\nonumber\\
&  -\;q\lambda^{2}f_{24}(L_{1}^{+})^{2}),\nonumber\\
(O_{f})_{L}^{14}  &  =\tilde{\Lambda}^{2}(f_{14}-q^{2}\lambda f_{24}L_{1}%
^{+}+\lambda f_{34}L_{2}^{+}),\nonumber\\
(O_{f})_{L}^{23}  &  =\tilde{\Lambda}^{2}(f_{23}-q\lambda f_{24}L_{1}%
^{+}+q\lambda f_{34}L_{2}^{+}),\nonumber\\
(O_{f})_{L}^{24}  &  =f_{24}\tilde{\Lambda}^{2}K_{1}^{-1},\nonumber\\
(O_{f})_{L}^{34}  &  =f_{34}\tilde{\Lambda}^{2}K_{2}^{-1},\nonumber\\%
[0.1in]%
(O_{f})_{L}^{123}  &  =\tilde{\Lambda}^{3}K_{1}^{1/2}K_{2}^{1/2}%
(f_{123}-q\lambda f_{124}L_{1}^{+}-q\lambda f_{134}L_{2}^{+}\\
&  +\;q^{2}\lambda^{2}f_{234}L_{1}^{+}L_{2}^{+}),\nonumber\\
(O_{f})_{L}^{124}  &  =\tilde{\Lambda}^{3}K_{1}^{-1/2}K_{2}^{1/2}%
(f_{124}-q\lambda f_{234}L_{2}^{+}),\nonumber\\
(O_{f})_{L}^{134}  &  =\tilde{\Lambda}^{3}K_{1}^{1/2}K_{2}^{-1/2}%
(f_{134}+q\lambda f_{234}L_{1}^{+}),\nonumber\\
(O_{f})_{L}^{234}  &  =f_{234}\tilde{\Lambda}^{3}K_{1}^{-1/2}K_{2}%
^{-1/2},\nonumber\\%
[0.1in]%
(O_{f})_{L}^{1234}  &  =f_{1234}\tilde{\Lambda}^{4},
\end{align}
and likewise for the other braiding
\begin{align}
(O_{f})_{\bar{L}}^{1}  &  =\tilde{\Lambda}^{-1}f_{1}K_{1}^{-1/2}K_{2}%
^{-1/2},\\
(O_{f})_{\bar{L}}^{2}  &  =\tilde{\Lambda}^{-1}K_{1}^{1/2}K_{2}^{-1/2}%
(f_{2}-q^{-1}\lambda f_{1}L_{1}^{-}),\nonumber\\
(O_{f})_{\bar{L}}^{3}  &  =\tilde{\Lambda}^{-1}K_{1}^{-1/2}K_{2}^{1/2}%
(f_{3}-q^{-1}\lambda f_{1}L_{2}^{-}),\nonumber\\
(O_{f})_{\bar{L}}^{4}  &  =\tilde{\Lambda}^{-1}K_{1}^{1/2}K_{2}^{1/2}%
(f_{4}+q^{-1}\lambda f_{3}L_{1}^{-}-q^{-1}\lambda f_{2}L_{2}^{-}\nonumber\\
&  -\;q^{-2}\lambda^{2}f_{1}L_{1}^{-}L_{2}^{-}),\nonumber\\%
[0.1in]%
(O_{f})_{\bar{L}}^{12}  &  =f_{12}\tilde{\Lambda}^{-2}K_{2}^{-1},\\
(O_{f})_{\bar{L}}^{13}  &  =f_{13}\tilde{\Lambda}^{-2}K_{1}^{-1},\nonumber\\
(O_{f})_{\bar{L}}^{14}  &  =\tilde{\Lambda}^{-2}(f_{14}+q^{-1}\lambda
f_{13}L_{1}^{-}+q^{-3}\lambda f_{12}L_{2}^{-}),\nonumber\\
(O_{f})_{\bar{L}}^{23}  &  =\tilde{\Lambda}^{-2}(f_{23}-q^{-2}\lambda
f_{13}L_{1}^{-}+q^{-2}\lambda f_{12}L_{2}^{-}),\nonumber\\
(O_{f})_{\bar{L}}^{24}  &  =\tilde{\Lambda}^{-2}K_{1}(f_{24}-\lambda
f_{14}L_{1}^{-}-q^{-1}\lambda f_{23}L_{1}^{-}\nonumber\\
&  -\;q^{-1}\lambda^{2}f_{13}(L_{1}^{-})^{2}),\nonumber\\
(O_{f})_{\bar{L}}^{34}  &  =\tilde{\Lambda}^{-2}K_{2}(f_{34}-\lambda
f_{14}L_{2}^{-}-q\lambda f_{23}L_{2}^{-}\nonumber\\
&  -\;q^{-1}\lambda^{2}f_{12}(L_{2}^{-})^{2}),\nonumber\\%
[0.1in]%
(O_{f})_{\bar{L}}^{123}  &  =f_{123}\tilde{\Lambda}^{-3}K_{1}^{-1/2}%
K_{2}^{-1/2},\\
(O_{f})_{\bar{L}}^{124}  &  =\tilde{\Lambda}^{-3}K_{1}^{1/2}K_{2}%
^{-1/2}(f_{124}+q^{-1}\lambda f_{123}L_{1}^{-}),\nonumber\\
(O_{f})_{\bar{L}}^{134}  &  =\tilde{\Lambda}^{-3}K_{1}^{-1/2}K_{2}%
^{1/2}(f_{134}-q^{-1}\lambda f_{123}L_{2}^{-}),\nonumber\\
(O_{f})_{\bar{L}}^{234}  &  =\tilde{\Lambda}^{-3}K_{1}^{1/2}K_{2}%
^{1/2}(f_{234}-q^{-1}\lambda f_{134}L_{1}^{-}+q^{-1}\lambda f_{124}L_{2}%
^{-}\nonumber\\
&  +\;q^{-2}\lambda^{2}f_{123}L_{1}^{-}L_{2}^{-}),\nonumber\\%
[0.1in]%
(O_{f})_{\bar{L}}^{1234}  &  =f_{1234}\tilde{\Lambda}^{-4}.
\end{align}

\section{q-Deformed Minkowski space\label{MinSpac}}

In this section we would like to focus on antisymmetrised q-Minkowski space
(its definition is given in Appendix \ref{AppQua}). If such a space is fused
together with its symmetrized counterpart, it gives a q-deformed superspace
useful for physical applications. Again, we start with the introduction of
Grassmann integrals:
\begin{align}
&  \int f(\theta^{-},\theta^{3/0},\theta^{3},\theta^{+})\;d_{L}^{4}\theta\\
&  \equiv-q^{-2}\partial_{\theta}^{-}\partial_{\theta}^{0}\partial_{\theta
}^{3/0}\partial_{\theta}^{+}\rhd f(\theta^{-},\theta^{3/0},\theta^{3}%
,\theta^{+})=f_{-,3/0,3+},\nonumber\\%
[0.1in]%
&  \int d_{R}^{4}\theta\;f(\theta^{-},\theta^{3/0},\theta^{3},\theta^{+})\\
&  \equiv-q^{2}f(\theta^{+},\theta^{3},\theta^{3/0},\theta^{-})\lhd
\hat{\partial}_{\theta}^{+}\hat{\partial}_{\theta}^{3/0}\hat{\partial}%
_{\theta}^{0}\hat{\partial}_{\theta}^{-}=f_{+3,3/0,-}.\nonumber
\end{align}
A direct calculation using the explicit form for the action of partial
derivatives on supernumbers \cite{Mik04}, shows for left superintegrals that
\begin{gather}
\int\theta^{-}\theta^{3/0}\theta^{3}\theta^{+}\,d_{L}^{4}\theta=1,\\
\int\theta^{\mu}\;d_{L}^{4}\theta=0,\quad\mu\in{\{}+,3/0,3,-{\}},\nonumber\\
\int\theta^{\nu}\theta^{\rho}\;d_{L}^{4}\theta=0,\quad\int\theta^{\alpha
}\theta^{\beta}\theta^{\gamma}\;d_{L}^{4}\theta=0,\nonumber
\end{gather}
with
\begin{align}
(\nu,\rho)  &  \in\{(-,+),(-,3/0),(-,3),(3,+),(3,3/0),(3/0,+)\},\\
(\alpha,\beta,\gamma)  &  \in
\{(-,3/0,3),(-,3/0,+),(-,3,+),(3/0,3,+)\}.\nonumber
\end{align}
and likewise for right superintegrals,%
\begin{gather}
\int d_{R}^{4}\theta\;\theta^{+}\theta^{3}\theta^{3/0}\theta^{-}=1,\\
\int d_{R}^{4}\theta\;\theta^{\mu}=0,\quad\mu\in{\{}+,3,3/0,-{\}},\nonumber\\
\int d_{R}^{4}\theta\;\theta^{\nu}\theta^{\rho}=0,\quad\int d_{R}^{4}%
\theta\;\theta^{\alpha}\theta^{\beta}\theta^{\gamma}=0,\nonumber
\end{gather}
where%
\begin{align}
(\nu,\rho)  &  \in\{(+,-),(3/0,-),(3,-),(+,3),(3,3/0),(+,3/0)\},\\
(\alpha,\beta,\gamma)  &  \in
\{(3,3/0,-),(+,3/0,-),(+,3,-),(+,3,3/0)\}.\nonumber
\end{align}
In the same way we can prove translational invariance given by
\begin{align}
\int\partial_{\theta}^{\mu}\rhd f(\theta^{-},\theta^{3/0},\theta^{3}%
,\theta^{+})\;d_{L}^{4}\theta &  =0,\\
\int d_{R}^{4}\theta\;f(\theta^{+},\theta^{3},\theta^{3/0},\theta^{-})\lhd
\hat{\partial}_{\theta}^{\mu}  &  =0.\nonumber
\end{align}
By performing the substitutions
\begin{gather}
d_{L}^{4}\theta\leftrightarrow d_{\bar{L}}^{4}\theta,\quad d_{R}^{4}%
\theta\leftrightarrow d_{\bar{R}}^{4}\theta,\\
f(\theta^{-},\theta^{3},\theta^{3/0},\theta^{+})\leftrightarrow f(\theta
^{+},\theta^{3},\theta^{3/0},\theta^{-}),\nonumber\\
f(\theta^{+},\theta^{3/0},\theta^{3},\theta^{-})\leftrightarrow f(\theta
^{-},\theta^{3/0},\theta^{3},\theta^{+}),\nonumber\\
\theta^{\pm}\leftrightarrow\theta^{\mp},\quad(\partial_{\theta})^{\pm
}\leftrightarrow(\hat{\partial}_{\theta})^{\mp},\quad q\leftrightarrow
q^{-1},\nonumber\\
\rhd\leftrightarrow\bar{\rhd},\quad\lhd\leftrightarrow\bar{\lhd},\nonumber
\end{gather}
we get the corresponding expressions for the conjugated differential calculus.

It is not very difficult to find out that the two bases described through the
pairings below are dual to each other:%
\begin{gather}
\langle\partial_{\theta}^{+},\theta^{-}\rangle_{L,\bar{R}}=-q,\quad
\langle\partial_{\theta}^{3/0},\theta^{3}\rangle_{L,\bar{R}}=1,\\
\langle\partial_{\theta}^{0},\theta^{3/0}\rangle_{L,\bar{R}}=1,\quad
\langle\partial_{\theta}^{-},\theta^{+}\rangle_{L,\bar{R}}=-q^{-1},\nonumber\\%
[0.1in]%
\langle\partial_{\theta}^{-}\partial_{\theta}^{+},\theta^{-}\theta^{+}%
\rangle_{L,\bar{R}}=1,\quad\langle\partial_{\theta}^{3/0}\partial_{\theta}%
^{+},\theta^{-}\theta^{3}\rangle_{L,\bar{R}}=-q,\\
\langle\partial_{\theta}^{-}\partial_{\theta}^{3/0},\theta^{3}\theta
^{+}\rangle_{L,\bar{R}}=-q^{-1},\quad\langle\partial_{\theta}^{0}%
\partial_{\theta}^{3/0},\theta^{3/0}\theta^{3}\rangle_{L,\bar{R}}%
=-q^{-2},\nonumber\\
\langle\partial_{\theta}^{0}\partial_{\theta}^{+},\theta^{-}\theta^{0}%
\rangle_{L,\bar{R}}=2q^{2}\lambda_{+}^{-1},\quad\langle\partial_{\theta}%
^{-}\partial_{\theta}^{0},\theta^{0}\theta^{+}\rangle_{L,\bar{R}}=2\lambda
_{+}^{-1},\nonumber\\%
[0.1in]%
\langle\partial_{\theta}^{0}\partial_{\theta}^{3/0}\partial_{\theta}%
^{+},\theta^{-}\theta^{3/0}\theta^{3}\rangle_{L,\bar{R}}=q^{3},\quad
\langle\partial_{\theta}^{-}\partial_{\theta}^{3/0}\partial_{\theta}%
^{+},\theta^{-}\theta^{3}\theta^{+}\rangle_{L,\bar{R}}=1,\\
\langle\partial_{\theta}^{-}\partial_{\theta}^{0}\partial_{\theta}%
^{3/0},\theta^{3/0}\theta^{3}\theta^{+}\rangle_{L,\bar{R}}=q,\quad
\langle\partial_{\theta}^{-}\partial_{\theta}^{0}\partial_{\theta}^{+}%
,\theta^{-}\theta^{0}\theta^{+}\rangle_{L,\bar{R}}=-2q\lambda_{+}%
^{-1},\nonumber\\%
[0.1in]%
\langle\partial_{\theta}^{-}\partial_{\theta}^{3/0}\partial_{\theta}%
^{3}\partial_{\theta}^{+},\theta^{-}\theta^{0}\theta^{3/0}\theta^{+}%
\rangle_{L,\bar{R}}=-q^{2}.
\end{gather}
By virtue of these identities, the exponential is given by
\begin{align}
&  \mbox{exp}(\theta_{\bar{R}}\mid(\partial_{\theta})_{L})\\
&  =1\otimes1-q^{-1}\theta^{-}\otimes\partial_{\theta}^{+}+\theta^{3}%
\otimes\partial_{\theta}^{3/0}+\theta^{3/0}\otimes\partial_{\theta}%
^{0}\nonumber\\
&  -\;q\theta^{+}\otimes\partial_{\theta}^{-}+\theta^{-}\theta^{+}%
\otimes\partial_{\theta}^{-}\partial_{\theta}^{+}-q^{-1}\theta^{-}\theta
^{3}\otimes\partial_{\theta}^{3/0}\partial_{\theta}^{+}\nonumber\\
&  -\;q\theta^{3}\theta^{+}\otimes\partial_{\theta}^{-}\partial_{\theta}%
^{3/0}-q^{-2}\theta^{3/0}\theta^{3}\otimes\partial_{\theta}^{0}\partial
_{\theta}^{3/0}\nonumber\\
&  +\;\frac{1}{2}q^{-2}\lambda_{+}\theta^{-}\theta^{0}\otimes\partial_{\theta
}^{0}\partial_{\theta}^{+}+\frac{1}{2}\lambda_{+}\theta^{0}\theta^{+}%
\otimes\partial_{\theta}^{-}\partial_{\theta}^{0}\nonumber\\
&  +\;q^{-3}\theta^{-}\theta^{3/0}\theta^{3}\otimes\partial_{\theta}%
^{0}\partial_{\theta}^{3/0}\partial_{\theta}^{+}+\theta^{-}\theta^{3}%
\theta^{+}\otimes\partial_{\theta}^{-}\partial_{\theta}^{3/0}\partial_{\theta
}^{+}\nonumber\\
&  +\;q^{-1}\theta^{3/0}\theta^{3}\theta^{+}\otimes\partial_{\theta}%
^{-}\partial_{\theta}^{0}\partial_{\theta}^{3/0}-\frac{1}{2}q^{-1}\lambda
_{+}\theta^{-}\theta^{0}\theta^{+}\otimes\partial_{\theta}^{-}\partial
_{\theta}^{0}\partial_{\theta}^{+}\nonumber\\
&  -\;q^{2}\theta^{-}\theta^{3/0}\theta^{3}\theta^{+}\otimes\partial_{\theta
}^{-}\partial_{\theta}^{0}\partial_{\theta}^{3/0}\partial_{\theta}%
^{+}.\nonumber
\end{align}
Furthermore, we have the crossing symmetries through
\begin{align}
\langle\underline{\hat{\partial}}_{\theta},\underline{\theta}\rangle_{\bar
{L},R}  &  \overset{%
\genfrac{}{}{0pt}{}{+\leftrightarrow-}{q\leftrightarrow1/q}%
}{\longleftrightarrow}\langle\underline{{\partial}}_{\theta},{\theta}%
\rangle_{{L},\bar{R}}\;,\\
\langle\underline{\theta},\underline{\hat{\partial}}_{\theta}\rangle_{\bar
{L},R}  &  \overset{%
\genfrac{}{}{0pt}{}{+\leftrightarrow-}{q\leftrightarrow1/q}%
}{\longleftrightarrow}\langle\underline{\theta},\underline{\partial}_{\theta
}\rangle_{{L},\bar{R}}\;,\nonumber\\%
[0.1in]%
\langle\underline{\hat{\partial}}_{\theta},\underline{\theta}\rangle_{\bar
{L},R}  &  \overset{+\leftrightarrow-}{\longleftrightarrow}\langle
\underline{\theta},\underline{\hat{\partial}}_{\theta}\rangle_{\bar{L},R}\;,\\
\langle\underline{{\partial}}_{\theta},\underline{\theta}\rangle_{{L},\bar
{R}}  &  \overset{+\leftrightarrow-}{\longleftrightarrow}\langle
\underline{\theta},\underline{{\partial}}_{\theta}\rangle_{{L},\bar{R}%
}\;,\nonumber
\end{align}
and
\begin{align}
\mbox{exp}(\theta_{{R}}\mid(\hat{\partial}_{\theta})_{\bar{L}})  &  \overset{%
\genfrac{}{}{0pt}{}{+\leftrightarrow-}{q\leftrightarrow1/q}%
}{\longleftrightarrow}\mbox{exp}(\theta_{{\bar{R}}}\mid(\partial_{\theta}%
)_{L})\;,\\
\mbox{exp}((\hat{\partial}_{\theta})_{{R}}\mid\theta_{\bar{L}})  &  \overset{%
\genfrac{}{}{0pt}{}{+\leftrightarrow-}{q\leftrightarrow1/q}%
}{\longleftrightarrow}\mbox{exp}((\partial_{\theta})_{\bar{R}}\mid\theta
_{L})\;,\nonumber\\%
[0.1in]%
\mbox{exp}(\theta_{R}\mid(\hat{\partial}_{\theta})_{\bar{L}})  &
\overset{+\leftrightarrow-}{\longleftrightarrow}\mbox{exp}((\hat{\partial
}_{\theta})_{R}\mid\theta_{\bar{L}})\;,\\
\mbox{exp}(\theta_{\bar{R}}\mid(\partial_{\theta})_{{L}})  &  \overset
{+\leftrightarrow-}{\longleftrightarrow}\mbox{exp}((\partial_{\theta}%
)_{\bar{R}}\mid\theta_{L})\;,\nonumber
\end{align}
where the transition symbols have the very same meaning as in Sec. \ref{Sec3}.

Next we would like to provide formulae for the coproduct of Grassmann
variables. On a basis of normal ordered monomials we have found
\begin{align}
\Delta_{L}(\theta^{\mu})  &  =\theta_{l}^{\mu}+\theta_{r}^{\mu},\quad\mu
\in{\{}+,3/0,0,-,{\},}\\%
[0.15in]%
\Delta_{L}(\theta^{+}\theta^{3/0})  &  =\theta_{l}^{+}\theta_{l}^{3/0}%
+\theta_{l}^{+}\theta_{r}^{3/0}-q^{-2}\theta_{l}^{3/0}\theta_{r}^{+}%
+\theta_{r}^{+}\theta_{r}^{3/0},\\
\Delta_{L}(\theta^{+}\theta^{0})  &  =\theta_{l}^{+}\theta_{l}^{0}+\theta
_{l}^{+}\theta_{r}^{0}-\theta_{l}^{0}\theta_{r}^{+}+\theta_{r}^{+}\theta
_{r}^{0}\nonumber\\
&  +\;\lambda\lambda_{+}^{-1}\theta_{l}^{+}\theta_{r}^{3/0}-q^{-2}%
\lambda\lambda_{+}^{-1}\theta_{l}^{3/0}\theta_{r}^{+},\nonumber\\
\Delta_{L}(\theta^{+}\theta^{-})  &  =\theta_{l}^{+}\theta_{l}^{-}+\theta
_{l}^{+}\theta_{r}^{-}-q^{-2}\theta_{l}^{-}\theta_{r}^{+}+\theta_{r}^{+}%
\theta_{r}^{-}\nonumber\\
&  +\;q^{-1}\lambda\lambda_{+}^{-1}\theta_{l}^{3/0}\theta_{r}^{3/0}%
,\nonumber\\
\Delta_{L}(\theta^{3/0}\theta^{0})  &  =\theta_{l}^{3/0}\theta_{l}^{0}%
+\theta_{l}^{3/0}\theta_{r}^{0}-q^{-2}\theta_{l}^{0}\theta_{r}^{3/0}%
+\theta_{r}^{3/0}\theta_{r}^{0}\nonumber\\
&  +\;\lambda\lambda_{+}^{-1}\theta_{l}^{3/0}\theta_{r}^{3/0}-q^{-2}%
\lambda\theta_{l}^{-}\theta_{r}^{+},\nonumber\\
\Delta_{L}(\theta^{3/0}\theta^{-})  &  =\theta_{l}^{3/0}\theta_{l}^{-}%
+\theta_{l}^{3/0}\theta_{r}^{-}-q^{-2}\theta_{l}^{-}\theta_{r}^{3/0}%
+\theta_{r}^{3/0}\theta_{r}^{-},\nonumber\\
\Delta_{L}(\theta^{0}\theta^{-})  &  =\theta_{l}^{0}\theta_{l}^{-}+\theta
_{l}^{0}\theta_{r}^{-}-\theta_{l}^{-}\theta_{r}^{0}+\theta_{r}^{0}\theta
_{r}^{-}\nonumber\\
&  +\;\lambda\lambda_{+}^{-1}\theta_{l}^{3/0}\theta_{r}^{-}-q^{-2}%
\lambda\lambda_{+}^{-1}\theta_{l}^{-}\theta_{r}^{3/0},\nonumber\\%
[0.15in]%
\Delta_{L}(\theta^{+}\theta^{3/0}\theta^{0})  &  =\theta_{l}^{+}\theta
_{l}^{3/0}\theta_{l}^{0}+\theta_{l}^{+}\theta_{l}^{3/0}\theta_{r}^{0}%
-q^{-2}\theta_{l}^{+}\theta_{l}^{0}\theta_{r}^{3/0}\\
&  +\;q^{-2}\theta_{l}^{3/0}\theta_{l}^{0}\theta_{r}^{+}+\theta_{l}^{+}%
\theta_{r}^{3/0}\theta_{r}^{0}-q^{-2}\theta_{l}^{3/0}\theta_{r}^{+}\theta
_{r}^{0}+q^{-2}\theta_{l}^{0}\theta_{r}^{+}\theta_{r}^{3/0}\nonumber\\
&  +\;\theta_{r}^{+}\theta_{r}^{3/0}\theta_{r}^{0}+q^{-1}\lambda\theta_{l}%
^{+}\theta_{l}^{3/0}\theta_{r}^{3/0}-q^{-2}\lambda\theta_{l}^{+}\theta_{l}%
^{-}\theta_{r}^{+},\nonumber\\
\Delta_{L}(\theta^{+}\theta^{3/0}\theta^{-})  &  =\theta_{l}^{+}\theta
_{l}^{3/0}\theta_{l}^{-}+\theta_{l}^{+}\theta_{l}^{3/0}\theta_{r}^{-}%
-q^{-2}\theta_{l}^{+}\theta_{l}^{-}\theta_{r}^{3/0}+q^{-4}\theta_{l}%
^{3/0}\theta_{l}^{-}\theta_{r}^{+}\nonumber\\
&  +\;\theta_{l}^{+}\theta_{r}^{3/0}\theta_{r}^{-}-q^{-2}\theta_{l}%
^{3/0}\theta_{r}^{+}\theta_{r}^{-}+q^{-4}\theta_{l}^{-}\theta_{r}^{+}%
\theta_{r}^{3/0}+\theta_{r}^{+}\theta_{r}^{3/0}\theta_{r}^{-},\nonumber\\
\Delta_{L}(\theta^{+}\theta^{0}\theta^{-})  &  =\theta_{l}^{+}\theta_{l}%
^{0}\theta_{l}^{-}+\theta_{l}^{+}\theta_{l}^{0}\theta_{r}^{-}-\theta_{l}%
^{+}\theta_{l}^{-}\theta_{r}^{0}+q^{-2}\theta_{l}^{0}\theta_{l}^{-}\theta
_{r}^{+}\nonumber\\
&  +\;\theta_{l}^{+}\theta_{r}^{0}\theta_{r}^{-}-\theta_{l}^{0}\theta_{r}%
^{+}\theta_{r}^{-}+q^{-2}\theta_{l}^{-}\theta_{r}^{+}\theta_{r}^{0}+\theta
_{r}^{+}\theta_{r}^{0}\theta_{r}^{-}\nonumber\\
&  +\;\lambda\lambda_{+}^{-1}\theta_{l}^{+}\theta_{l}^{3/0}\theta_{r}%
^{-}+q^{-1}\lambda\lambda_{+}^{-1}\theta_{l}^{3/0}\theta_{l}^{0}\theta
_{r}^{3/0}\nonumber\\
&  +\;q^{-4}\lambda\lambda_{+}^{-1}\theta_{l}^{3/0}\theta_{l}^{-}\theta
_{r}^{+}+q\lambda(q\lambda-2)\theta_{l}^{+}\theta_{l}^{-}\theta_{r}%
^{3/0}\nonumber\\
&  +\;\lambda\lambda_{+}^{-1}\theta_{l}^{+}\theta_{r}^{3/0}\theta_{r}%
^{-}-2q^{-2}\lambda\lambda_{+}^{-1}\theta_{l}^{3/0}\theta_{r}^{+}\theta
_{r}^{-}\nonumber\\
&  -\;q^{-1}\lambda\lambda_{+}^{-1}\theta_{l}^{3/0}\theta_{r}^{3/0}\theta
_{r}^{0}+q^{-4}\lambda\lambda_{+}^{-1}\theta_{l}^{-}\theta_{r}^{+}\theta
_{r}^{3/0},\nonumber\\
\Delta_{L}(\theta^{3/0}\theta^{0}\theta^{-})  &  =\theta_{l}^{3/0}\theta
_{l}^{0}\theta_{l}^{-}+q^{-4}\theta_{l}^{0}\theta_{l}^{-}\theta_{r}%
^{3/0}-\theta_{l}^{3/0}\theta_{l}^{-}\theta_{r}^{0}\nonumber\\
&  -\;q^{-2}\theta_{l}^{0}\theta_{r}^{3/0}\theta_{r}^{+}+q^{-2}\theta_{l}%
^{-}\theta_{r}^{3/0}\theta_{r}^{0}+\theta_{r}^{3/0}\theta_{r}^{0}\theta
_{r}^{-}\nonumber\\
&  +\;q^{-3}\lambda\theta_{l}^{3/0}\theta_{l}^{-}\theta_{r}^{3/0}%
+q^{-2}\lambda\theta_{l}^{-}\theta_{r}^{+}\theta_{r}^{-},\nonumber\\%
[0.15in]%
\Delta_{L}(\theta^{+}\theta^{3/0}\theta^{0}\theta^{-})  &  =\theta_{l}%
^{+}\theta_{l}^{3/0}\theta_{l}^{0}\theta_{l}^{-}+\theta_{l}^{+}\theta
_{l}^{3/0}\theta_{l}^{0}\theta_{r}^{-}-\theta_{l}^{+}\theta_{l}^{3/0}%
\theta_{l}^{-}\theta_{r}^{0}\\
&  -\;q^{-4}\theta_{l}^{3/0}\theta_{l}^{0}\theta_{l}^{-}\theta_{r}^{+}%
+q^{-4}\theta_{l}^{+}\theta_{l}^{0}\theta_{l}^{-}\theta_{r}^{3/0}+\theta
_{l}^{+}\theta_{l}^{3/0}\theta_{r}^{0}\theta_{r}^{-}\nonumber\\
&  -\;q^{-2}\theta_{l}^{+}\theta_{l}^{0}\theta_{r}^{3/0}\theta_{r}^{-}%
+q^{-2}\theta_{l}^{+}\theta_{l}^{-}\theta_{r}^{3/0}\theta_{r}^{0}+q^{-2}%
\theta_{l}^{3/0}\theta_{l}^{0}\theta_{r}^{+}\theta_{r}^{-}\nonumber\\
&  -\;q^{-4}\theta_{l}^{3/0}\theta_{l}^{-}\theta_{r}^{+}\theta_{r}^{0}%
+q^{-6}\theta_{l}^{0}\theta_{l}^{-}\theta_{r}^{+}\theta_{r}^{3/0}+\theta
_{l}^{+}\theta_{r}^{3/0}\theta_{r}^{0}\theta_{r}^{-}\nonumber\\
&  -\;q^{-2}\theta_{l}^{3/0}\theta_{r}^{+}\theta_{r}^{0}\theta_{r}^{-}%
+q^{-2}\theta_{l}^{0}\theta_{r}^{+}\theta_{r}^{3/0}\theta_{r}^{-}+q^{-4}%
\theta_{l}^{-}\theta_{r}^{+}\theta_{r}^{3/0}\theta_{r}^{0}\nonumber\\
&  +\;\theta_{r}^{+}\theta_{r}^{3/0}\theta_{r}^{0}\theta_{r}^{-}-3q^{4}%
\lambda\lambda_{+}^{-1}\theta_{l}^{+}\theta_{l}^{3/0}\theta_{l}^{-}\theta
_{r}^{0}\nonumber\\
&  +\;q^{-2}\lambda\lambda_{+}^{-1}\theta_{l}^{+}\theta_{l}^{3/0}\theta
_{r}^{3/0}\theta_{r}^{-}-q^{-2}\lambda\theta_{l}^{+}\theta_{l}^{-}\theta
_{r}^{+}\theta_{r}^{-}\nonumber\\
&  -\;q^{6}\lambda\lambda_{+}^{-1}\theta_{l}^{3/0}\theta_{l}^{-}\theta_{r}%
^{+}\theta_{r}^{3/0}+q^{-2}\lambda\lambda_{+}^{-1}\theta_{l}^{3/0}\theta
_{r}^{+}\theta_{r}^{3/0}\theta_{r}^{-}.\nonumber
\end{align}
The expressions for the corresponding antipodes of our normal ordered
monomials are:
\begin{align}
&  S_{L}(\theta^{\mu})=-\theta^{\mu},\quad\mu\in\{+,3/0,0,-\},\\
&  S_{L}(\theta^{\nu}\theta^{\rho})=q^{-2}\theta^{\nu}\theta^{\rho
},\nonumber\\
&  S_{L}(\theta^{\alpha}\theta^{\beta}\theta^{\gamma})=-q^{-6}\theta^{\alpha
}\theta^{\beta}\theta^{\gamma},\nonumber\\
&  S_{L}(\theta^{+}\theta^{3/0}\theta^{0}\theta^{-})=q^{-12}\theta^{+}%
\theta^{3/0}\theta^{0}\theta^{-},\nonumber
\end{align}
where%
\begin{align}
(\nu,\rho)  &  \in\{(+,-),(3/0,-),(0,-),(+,0),(3/0,0),(+,3/0)\},\\
(\alpha,\beta,\gamma)  &  \in
\{(+,3/0,0),(+,3/0,-),(+,0,-),(3/0,0,-)\}.\nonumber
\end{align}
Notice that monomials with unspecified indices have to refer to the ordering
$\theta^{+}\theta^{3/0}\theta^{0}\theta^{-}$. In complete analogy to the
previous sections, one can check the crossing symmetries
\begin{align}
&  \Delta_{L},S_{L}\overset{%
\genfrac{}{}{0pt}{}{+\leftrightarrow-}{q\leftrightarrow1/q}%
}{\longleftrightarrow}\Delta_{\bar{L}},S_{\bar{L}},\\
&  \Delta_{R},S_{R}\overset{%
\genfrac{}{}{0pt}{}{+\leftrightarrow-}{q\leftrightarrow1/q}%
}{\longleftrightarrow}\Delta_{\bar{R}},S_{\bar{R}},\nonumber\\%
[0.10in]%
&  \Delta_{L},S_{L}\overset{%
\genfrac{}{}{0pt}{}{+\leftrightarrow-}{q\leftrightarrow1/q}%
}{\longleftrightarrow}\Delta_{R},S_{R},\\
&  \Delta_{\bar{L}},S_{\bar{L}}\overset{%
\genfrac{}{}{0pt}{}{+\leftrightarrow-}{q\leftrightarrow1/q}%
}{\longleftrightarrow}\Delta_{\bar{R}},S_{\bar{R}}.\nonumber
\end{align}

Finally, let us come to expressions for braided products concerning
supernumbers. Such braided products can be calculated from
\begin{align}
f(\theta^{+},\theta^{3/0},\theta^{0},\theta^{-})\,\underline{\odot}_{L/\bar
{L}}\,g  &  =g\otimes f^{\prime}+\sum\nolimits_{\underline{K}}\left(
(O_{f})_{L/\bar{L}}^{\underline{K}}\rhd g\right)  \otimes\theta^{\,\underline
{K}},\\
g\,\underline{\odot}_{R/\bar{R}}\,f(\theta^{+},\theta^{3/0},\theta^{0}%
,\theta^{-})  &  =f^{\prime}\otimes g+\sum\nolimits_{\underline{K}}%
\theta^{\,\underline{K}}\otimes\left(  g\lhd(O_{f})_{L/\bar{L}}^{\underline
{K}}\right)  .\nonumber
\end{align}
For brevity, we introduced the following combinations of symmetry generators
(for their action on quantum spaces see again Refs.\cite{BW01} and
\cite{Mik04}):%
\begin{align}
(O_{f})_{L}^{+}  &  =\tilde{\Lambda}(\tau^{3})^{-1/2}[f_{+}\,\sigma
^{2}-q^{1/2}\lambda\lambda_{+}^{1/2}f_{3/0}\,S^{1}\\
&  -\;\lambda^{2}f_{-}\,T^{-}S^{1}+q^{1/2}\lambda\lambda_{+}^{-1/2}%
f_{0}\,(T^{-}\sigma^{2}+qS^{1})],\nonumber\\
(O_{f})_{L}^{3/0}  &  =\tilde{\Lambda}[-q^{3/2}\lambda\lambda_{+}^{-1/2}%
f_{+}\,T^{2}+f_{3/0}\,\tau^{1}\nonumber\\
&  -\;\lambda_{+}^{-1}f_{0}\,(\lambda^{2}T^{-}T^{2}+q(\tau^{1}-\sigma
^{2}))\nonumber\\
&  +\;q^{-1/2}f_{-}\,\lambda\lambda_{+}^{-1/2}(\tau^{1}T^{-}-q^{-1}%
S^{1})],\nonumber\\
(O_{f})_{L}^{0}  &  =\tilde{\Lambda}(f_{0}\,\sigma^{2}-q^{-1/2}\lambda
\lambda_{+}^{1/2}f_{-}\,S^{1}),\nonumber\\
(O_{f})_{L}^{-}  &  =\tilde{\Lambda}(\tau^{3})^{1/2}(-q^{5/2}\lambda
\lambda_{+}^{-1/2}f_{0}\,T^{2}+f_{-}\,\tau^{1}),\nonumber\\%
[0.15in]%
(O_{f})_{L}^{+,3/0}  &  =\tilde{\Lambda}^{2}(\tau^{3})^{-1/2}[f_{+,3/0}%
+q^{-1}\lambda_{+}^{-1}f_{+0}\,((\sigma^{2})^{2}-1)\nonumber\\
&  +\;q^{-1/2}\lambda\lambda_{+}^{-1/2}f_{+-}\,(T^{-}-q^{-1}S^{1}\sigma
^{2})\nonumber\\
&  -\;q^{-1/2}\lambda\lambda_{+}^{-1/2}f_{3/0,0}\,(q^{-1}T^{-}+s^{1}\sigma
^{2})\nonumber\\
&  +\;q^{-1}\lambda^{2}f_{3/0,-}\,(S^{1})^{2}\nonumber\\
&  +\;\lambda^{2}\lambda_{+}f_{0-}\,((T^{-})^{2}-q^{-2}(S^{1})^{2}%
)],\nonumber\\
(O_{f})_{L}^{+0}  &  =\tilde{\Lambda}^{2}(\tau^{3})^{-1/2}[f_{+0}\,(\sigma
^{2})^{2}-q^{-1/2}\lambda\lambda_{+}^{1/2}f_{+-}\,S^{1}\sigma^{2})\nonumber\\
&  -\;q^{1/2}\lambda\lambda_{+}^{1/2}f_{3/0,0}\,S^{1}\sigma^{2}+\lambda
^{2}\lambda_{+}f_{3/0,-}\,(S^{1})^{2}\nonumber\\
&  -\;q^{-1}\lambda^{2}f_{0-}\,(S^{1})^{2}],\nonumber\\
(O_{f})_{L}^{+-}  &  =\tilde{\Lambda}^{2}[-q^{1/2}\lambda\lambda_{+}%
^{-1/2}f_{+0}\,T^{2}\sigma^{2}+f_{+-}\,(1+\lambda^{2}T^{2}S^{1})\nonumber\\
&  +\;q\lambda^{2}f_{3/0,0}\,T^{2}S^{1}-q^{-3/2}\lambda\lambda_{+}%
^{1/2}f_{3/0,-}\,\tau^{1}S^{1}\nonumber\\
&  +\;q^{1/2}\lambda\lambda_{+}^{-1/2}f_{0-}\,(T^{-}+q^{-3}\tau^{1}%
S^{1})],\nonumber\\
(O_{f})_{L}^{3/0,0}  &  =\tilde{\Lambda}^{2}[(-q^{3/2}\lambda\lambda
_{+}^{-1/2}f_{+0}\,T^{2}\sigma^{2}+q\lambda^{2}f_{+-}\,T^{2}S^{1})\nonumber\\
&  +\;f_{3/0,0}\,(1+q^{2}\lambda^{2}T^{2}S^{1})-q^{-1/2}\lambda\lambda
_{+}^{1/2}f_{3/0,-}\,\tau^{1}S^{1}\nonumber\\
&  +\;q^{-1/2}\lambda\lambda_{+}^{-1/2}f_{0-}\,(T^{-}+q^{-1}\tau^{1}%
S^{1})],\nonumber\\
(O_{f})_{L}^{3/0,-}  &  =\tilde{\Lambda}^{2}(\tau^{3})^{1/2}[q^{6}\lambda
^{2}\lambda_{+}^{-1}f_{+0}\,(T^{2})^{2}\nonumber\\
&  -\;q^{7/2}\lambda\lambda_{+}^{-1/2}f_{+-}\,T^{2}\tau^{1}+q^{9/2}%
\lambda\lambda_{+}^{-1/2}f_{3/0,0}\,T^{2}\tau^{1}\nonumber\\
&  +\;f_{3/0,-}\,(\tau^{1})^{2}+q^{-1}\lambda_{+}^{-1}f_{0-}\,(1-(\tau
^{1})^{2})],\nonumber\\
(O_{f})_{L}^{0-}  &  =\tilde{\Lambda}^{2}(\tau^{3})^{-1/2}f_{0-},\nonumber\\%
[0.15in]%
(O_{f})_{L}^{+,3/0,0}  &  =\tilde{\Lambda}^{3}(\tau^{3})^{-1/2}[f_{+,3/0,0}%
\,\sigma^{2}-q^{-1/2}\lambda\lambda_{+}^{-1/2}f_{+,3/0,-}\,S^{1}\nonumber\\
&  -\;q^{1/2}\lambda\lambda_{+}^{-1/2}f_{+0-}\,(qT^{-}\sigma^{2}+q^{-1}%
(q^{3}-\lambda_{+})S^{1})\nonumber\\
&  +\;\lambda^{2}f_{3/0,0,-}\,T^{-}S^{1}],\nonumber\\
(O_{f})_{L}^{+,3/0,-}  &  =\tilde{\Lambda}^{3}[-q^{5/2}\lambda\lambda
_{+}^{-1/2}f_{+,3/0,0}\,T^{2}+f_{+,3/0,-}\,\tau^{1}\nonumber\\
&  +\;\lambda_{+}^{-1}f_{+0-}\,(q^{2}\lambda^{2}T^{-}T^{2}+q^{-1}%
(1-q^{3}\lambda)(\sigma^{2}-\tau^{1}))\nonumber\\
&  +\;q^{-3/2}\lambda\lambda_{+}^{-1/2}f_{3/0,0,-}\,(T^{-}\tau^{1}%
+(q^{3}-\lambda_{+})S^{1})],\nonumber\\
(0_{f})_{L}^{+0-}  &  =\tilde{\Lambda}^{3}(f_{+0-}\,\sigma^{2}-q^{-3/2}%
\lambda\lambda_{+}^{1/2}f_{3/0,0,-}\,S^{1}),\nonumber\\
(O_{f})_{L}^{3/0,0-}  &  =\tilde{\Lambda}^{3}(\tau^{3})^{1/2}(-q^{7/2}%
\lambda\lambda_{+}^{-1/2}f_{+0-}\,T^{2}+f_{3/0,0,-}\,\tau^{1}),\nonumber\\%
[0.15in]%
(O_{f})_{L}^{+,3/0,0-}  &  =f_{+,3/0,0-}\tilde{\Lambda}^{4},
\end{align}
and
\begin{align}
(O_{f})_{\bar{L}}^{+}  &  =\tilde{\Lambda}^{-1}(f_{+}\,\sigma^{2}%
-q^{-1/2}\lambda\lambda_{+}^{-1/2}f_{0}\,S^{1}),\\
(O_{f})_{\bar{L}}^{3/0}  &  =\tilde{\Lambda}^{-1}(\tau^{3})^{-1/2}%
[-q^{1/2}\lambda\lambda_{+}^{-1/2}f_{+}\,(T^{+}\sigma^{2}+q\tau^{3}%
T^{2})\nonumber\\
&  +\;f_{3/0}\,\sigma^{2}+f_{0}\,(\lambda^{2}T^{+}S^{1}+q^{-1}(\tau^{3}%
\tau^{1}-\sigma^{2}))\nonumber\\
&  -\;q^{1/2}\lambda\lambda_{+}^{-1/2}f_{-}\,S^{1}],\nonumber\\
(O_{f})_{\bar{L}}^{0}  &  =\tilde{\Lambda}^{-1}(\tau^{3})^{1/2}(-q^{1/2}%
\lambda\lambda_{+}^{1/2}f_{+}\,T^{2}+f_{0}(\tau^{3})^{-1}\tau^{1}),\nonumber\\
(O_{f})_{\bar{L}}^{-}  &  =\tilde{\Lambda}^{-1}[q^{2}\lambda^{2}f_{+}%
\,T^{2}T^{+}-q^{3/2}\lambda\lambda_{+}^{1/2}f_{3/0}\,T^{2}\nonumber\\
&  -\;q^{1/2}\lambda\lambda_{+}^{-1/2}f_{0}\,(qT^{+}\tau^{1}-T^{2}%
))+f_{-}\,\tau^{1}],\nonumber\\%
[0.15in]%
(O_{f})_{\bar{L}}^{+,3/0}  &  =\tilde{\Lambda}^{-2}(\tau^{3})^{-1/2}%
[f_{+,3/0}\,(\sigma^{2})^{2}+q^{-1}\lambda_{+}^{-1}f_{+0}\,(\tau^{3}%
-(\sigma^{2})^{2})\nonumber\\
&  -\;q^{1/2}\lambda\lambda_{+}^{-1/2}f_{+-}\,S^{1}\sigma^{2}+q^{-1/2}%
\lambda\lambda_{+}^{-1/2}f_{3/0,0}\,S^{1}\sigma^{2}\nonumber\\
&  +\;q^{2}\lambda^{2}\lambda_{+}^{-1}f_{0-}\,(S^{1})^{2})],\nonumber\\
(O_{f})_{\bar{L}}^{+0}  &  =\tilde{\Lambda}^{-2}(\tau^{3})^{1/2}%
f_{+0}\,,\nonumber\\
(O_{f})_{\bar{L}}^{+-}  &  =\tilde{\Lambda}^{-2}[-q^{3/2}\lambda\lambda
_{+}^{1/2}f_{+,3/0}\,T^{2}\sigma^{2}\nonumber\\
&  +\;q^{1/2}\lambda\lambda_{+}^{-1/2}f_{+0}\,(T^{2}\sigma^{2}-qT^{+}%
)+f_{+-}(1+q^{2}\lambda^{2}T^{2}S^{1})\nonumber\\
&  -\;q\lambda^{2}f_{3/0,0}\,T^{2}S^{1}-q^{3/2}\lambda\lambda_{+}^{-1/2}%
f_{0-}\,\tau^{1}S^{1}],\nonumber\\
(O_{f})_{\bar{L}}^{3/0,0}  &  =\tilde{\Lambda}^{-2}[q^{1/2}\lambda\lambda
_{+}^{1/2}f_{+,3/0}\,T^{2}\sigma^{2}\nonumber\\
&  -\;q\lambda^{2}f_{+-}\,T^{2}S^{1}+f_{3/0,0}\,(1+\lambda^{2}T^{2}%
S^{1})\nonumber\\
&  +\;\lambda\lambda{+}^{-1/2}f_{+0}\,(q^{1/2}\,\tau^{1}S^{1}-q^{-1/2}%
\,(T^{2}\sigma^{2}+q^{3}T^{+}))],\nonumber\\
(O_{f})_{\bar{L}}^{3/0,-}  &  =\tilde{\Lambda}^{-2}(\tau^{3})^{1/2}%
[q^{3}\lambda^{2}f_{+,3/0}\,(T^{2})^{2}\nonumber\\
&  +\;q^{2}\lambda^{2}\lambda_{+}^{-1}f_{+0}\,((\tau^{3})^{-1}(T^{+}%
)^{2}-(T^{2})^{2})\nonumber\\
&  -\;q^{1/2}\lambda\lambda_{+}^{-1/2}f_{+-}\,((\tau^{3})^{-1}T^{+}+qT^{2}%
\tau^{1})\nonumber\\
&  +\;q^{1/2}\lambda\lambda_{+}^{-1/2}f_{3/0,0}\,(T^{2}\tau^{1}-q(\tau
^{3})^{-1}T^{+})\nonumber\\
&  -\;q^{-1}\lambda_{+}^{-1}f_{0-}\,(1-(\tau^{1})^{2})],\nonumber\\
(O_{f})_{\bar{L}}^{0-}  &  =\tilde{\Lambda}^{-2}(\tau^{3})^{1/2}[q^{4}%
\lambda^{2}\lambda_{+}f_{+,3/0}\,(T^{2})^{2}-q^{3}\lambda^{2}f_{+0}%
\,(T^{2})^{2}\nonumber\\
&  -\;q^{5/2}\lambda\lambda{+}^{1/2}f_{+-}\,T^{2}\tau^{1}+q^{3/2}%
\lambda\lambda{+}^{1/2}f_{3/0,0}\,T^{2}\tau^{1}+f_{0-}\,(\tau^{1}%
)^{2}],\nonumber\\%
[0.15in]%
(O_{f})_{\bar{L}}^{+,3/0,0}  &  =\tilde{\Lambda}^{-3}(f_{+,3/0,0}\,\sigma
^{2}+q^{1/2}\lambda\lambda_{+}^{-1/2}f_{+0-}\,S^{1})\nonumber\\
(O_{f})_{\bar{L}}^{+,3/0,-}  &  =\tilde{\Lambda}^{-3}(\tau^{3})^{1/2}%
[q^{1/2}\lambda\lambda_{+}^{1/2}f_{+,3/0,0}\,(T^{2}-q(\tau^{3})^{-1}%
T^{+}\sigma^{2})\nonumber\\
&  +f_{+,3/0,-}\,(\tau^{3})^{-1}\sigma^{2}+q^{3/2}\lambda\lambda_{+}%
^{-1/2}f_{3/0,0-}\,(\tau^{3})^{-1}S^{1}\nonumber\\
&  -\lambda_{+}^{-1}f_{+0-}\,(q^{-1}(\tau^{3})^{-1}\sigma^{2}-\tau^{1}%
+q^{2}\lambda^{2}(\tau^{3})^{-1}T^{+}S^{1})],\nonumber\\
(O_{f})_{\bar{L}}^{+0-}  &  =\tilde{\Lambda}^{-3}(\tau^{3})^{1/2}%
(q^{3/2}\lambda\lambda_{+}^{1/2}f_{+,3/0,0}\,T^{2}+f_{+0-}\,\tau
^{1}),\nonumber\\
(O_{f})_{\bar{L}}^{3/0,0-}  &  =\tilde{\Lambda}^{-3}[-q^{4}\lambda
^{2}f_{+,3/0,0}\,T^{+}T^{2}+q^{1/2}\lambda\lambda_{+}^{1/2}f_{+,3/0,-}%
\,T^{2}\nonumber\\
&  -\;q^{1/2}\lambda\lambda_{+}^{-1/2}f_{+0-}\,(q^{-1}T^{2}+q^{2}T^{+}\tau
^{1})+f_{3/0,0,-}\,\tau^{1}],\nonumber\\%
[0.15in]%
(O_{f})_{\bar{L}}^{+,3/0,0-}  &  =\tilde{\Lambda}^{-4}f_{+,3/0,0-}.
\end{align}

\section{Conclusion\label{AppA}}

Let us end with a few comments on our results. In the last four sections we
have provided q-analogs for elements of superanalysis. In doing so we have
realized that on q-deformed quantum spaces Grassmann integrals, Grassmann
exponentials and Grassmann translations can be constructed in complete analogy
to the classical case. That this analogy is a really far reaching one can also
be seen from the fact that translational invariance of our integrals implies
rules for integration by parts:
\begin{align}
\int f(\underline{\theta})\left[  (\partial_{\theta})_{A}\rhd g(\underline
{\theta})\right]  d_{L}^{n}\theta &  =\int\left[  f(\underline{\theta}%
)\lhd(\partial_{\theta})_{A}\right]  g(\underline{\theta})\,d_{L}^{n}\theta,\\
\int d_{R}^{n}\theta\left[  f(\underline{\theta})\lhd(\partial_{\theta}%
)_{A}\right]  g(\underline{\theta})  &  =\int d_{R}^{n}\theta\,f(\underline
{\theta})\left[  (\partial_{\theta})_{A}\rhd g(\underline{\theta})\right]
.\nonumber
\end{align}
For this to verify, one has to take into account that
\begin{align}
f(\partial\rhd g)  &  =\partial_{(2)}\rhd{[}(f\lhd\partial_{(1)})g{]},\\
(f\lhd\partial)g  &  ={[}f(\partial_{(1)}\rhd g){]}\lhd\partial_{(2)}%
.\nonumber
\end{align}
Furthermore, it should be stressed that translational invariance of our
Grassmann integrals can alternatively be expressed as
\begin{align}
&  (1\otimes\int\;.\;d_{L}^{n}\theta)\circ\Delta_{\bar{R}}f(\underline{\theta
})=(\int\;.\;d_{L}^{n}\theta\otimes1)\circ\Delta_{\bar{R}}f(\underline{\theta
})\\
&  =\int f(\underline{\theta})\,d_{L}^{n}\theta,\nonumber\\%
[0.1in]%
&  (1\otimes\int d_{R}^{n}\theta\;.\;)\circ\Delta_{L}f(\underline{\theta
})=(\int d_{R}^{n}\theta\;.\;\otimes1)\circ\Delta_{L}f(\underline{\theta})\\
&  =\int d_{R}^{n}\theta\,f(\underline{\theta}).\nonumber
\end{align}
The above statements can be proved in a straightforward manner by insertion of
the explicit expressions for superintegral and coproduct. Let us also notice
that in Ref. \cite{Maj95} this property was taken as abstract definition for
an integral on quantum spaces. In our case, integrals are given by explicit
instructions being compatible with the requirement of translational invariance.

Next, let us make contact with q-analogs of $\delta$-functions. For a $\delta
$-function on q-deformed Grassmann algebras we require to hold:
\begin{align}
\int f(\underline{\theta})\,\delta_{\bar{L}/L}^{n}(\underline{\theta
})\,d_{\bar{L}/L}^{n}\theta &  =\int\delta_{\bar{L}/L}^{n}(\underline{\theta
})\,f(\underline{\theta})\,d_{\bar{L}/L}^{n}\theta=f^{\prime},\nonumber\\
\int d_{\bar{R}/R}^{n}\theta\,f(\underline{\theta})\,\delta_{\bar{R}/R}%
^{n}\,(\underline{\theta})  &  =\int d_{\bar{R}/R}^{n}\theta\,\delta_{\bar
{R}/R}^{n}(\underline{\theta})\,f(\underline{\theta})=f^{\prime}.
\end{align}
It is not very difficult to show that these requirements are satisfied by

\begin{enumerate}
\item[a)] (quantum plane)\newline%
\begin{align}
\delta_{L}^{2}(\underline{\theta})  &  =\delta_{\bar{R}}^{2}(\underline
{\theta})=\theta^{2}\theta^{1},\\
\delta_{\bar{L}}^{2}(\underline{\theta})  &  =\delta_{{R}}^{2}(\underline
{\theta})=\theta^{1}\theta^{2},\nonumber
\end{align}

\item[b)] (three-dimensional Euclidean space)\newline%
\begin{align}
\delta_{L}^{3}(\underline{\theta})  &  =\delta_{\bar{R}}^{3}(\underline
{\theta})=\theta^{+}\theta^{3}\theta^{-},\\
\delta_{\bar{L}}^{3}(\underline{\theta})  &  =\delta_{{R}}^{3}(\underline
{\theta})=\theta^{-}\theta^{3}\theta^{+},\nonumber
\end{align}

\item[c)] (four-dimensional Euclidean space)\newline%
\begin{align}
\delta_{L}^{4}(\underline{\theta})  &  =\delta_{\bar{R}}^{4}(\underline
{\theta})=\theta^{4}\theta^{3}\theta^{2}\theta^{1},\\
\delta_{\bar{L}}^{4}(\underline{\theta})  &  =\delta_{{R}}^{4}(\underline
{\theta})=\theta^{1}\theta^{2}\theta^{3}\theta^{4},\nonumber
\end{align}

\item[d)] (q-deformed Minkowski space)\newline%
\begin{align}
\delta_{L}^{4}(\underline{\theta})  &  =\theta^{-}\theta^{3/0}\theta^{3}%
\theta^{+},\quad\delta_{R}^{4}(\underline{\theta})=\theta^{+}\theta^{3}%
\theta^{3/0}\theta^{-},\\
\delta_{\bar{L}}^{4}(\underline{\theta})  &  =\theta^{+}\theta^{3/0}\theta
^{3}\theta^{-},\quad\delta_{\bar{R}}^{4}(\underline{\theta})=\theta^{-}%
\theta^{3}\theta^{3/0}\theta^{+}.\nonumber
\end{align}

\end{enumerate}

Last but not least we would like to say a few words about the connection
between q-deformed superexponentials and translations of q-deformed
supernumbers. That translations on quantum spaces are indeed given by the
coproduct can also be seen from the existence of some sort of q-deformed
Taylor rules for which we have \cite{Maj-93/2}%
\begin{align}
f(\psi\,\underline{\oplus}_{L}\,\theta)  &  =\exp(\psi_{R}\mid(\hat{\partial
}_{\theta})_{\bar{L}})\,\bar{\triangleright}\,f(\underline{\theta
}),\label{MajForExp}\\
f(\psi\,\underline{\oplus}_{\bar{L}}\,\theta)  &  =\exp(\psi_{\bar{R}}%
\mid(\partial_{\theta})_{L})\triangleright f(\underline{\theta}),\nonumber\\%
[0.10in]%
f(\theta\,\underline{\oplus}_{\bar{R}}\,\psi)  &  =f(\underline{\theta
})\triangleleft\exp((\hat{\partial}_{\theta})_{R}\mid\psi_{\bar{L}}),\\
f(\theta\,\underline{\oplus}_{R}\,\psi)  &  =f(\underline{\theta}%
)\,\bar{\triangleleft}\,\exp((\partial_{\theta})_{\bar{R}}\mid\psi
_{L}).\nonumber
\end{align}
Again, these identities can be verified in a straightforward manner making use
of the explicit form for the superexponentials and the action of derivatives
on antisymmetrized quantum spaces.\vspace{0.2cm}

\noindent\textbf{Acknowledgements}\newline First of all we are very grateful
to Eberhard Zeidler for his invitation to the MPI Leipzig, his interesting and
useful discussions, his special interest in our work and his financial
support. Also we want to express our gratitude to Julius Wess for his efforts
and his steady support. Furthermore we would like to thank Fabian Bachmaier
for teaching us Mathematica and Florian Koch for useful suggestions. Finally,
we thank Dieter L\"{u}st for kind hospitality.

\appendix

\section{Quantum spaces\label{AppQua}}

In this appendix we list for the quantum spaces under consideration the
explicit form of their defining relations and the non-vanishing elements of
their quantum metrics.

The coordinates of two-dimensional antisymmetrized Manin plane fulfill the
relation \cite{Man88,SS90}
\begin{equation}
\theta^{1}\theta^{2}=-q^{-1}\theta^{2}\theta^{1}.
\end{equation}
The q-deformed spinor metric is given by a matrix $\varepsilon^{ij}$ with
non-vanishing elements
\begin{equation}
\varepsilon^{12}=q^{-1/2},\quad\varepsilon^{21}=-q^{1/2}.
\end{equation}
Indices can be raised and lowered as usual, i.e.
\begin{equation}
\theta^{\alpha}=\varepsilon^{\alpha\beta}\theta_{\beta},\quad\theta_{\alpha
}=\varepsilon_{\alpha\beta}\theta^{\beta},
\end{equation}
where $\varepsilon_{ij}=-\varepsilon^{ij}.$

The commutation relations defining an antisymmetrized version of
three-dimensional q-deformed Euclidean space read
\begin{align}
(\theta^{+})^{2}  &  =(\theta^{-})^{2}=0,\\
(\theta^{3})^{2}  &  =\lambda\theta^{+}\theta^{-},\nonumber\\
\theta^{+}\theta^{-}  &  =-\theta^{-}\theta^{+},\nonumber\\
\theta^{\pm}\theta^{3}  &  =-q^{\pm2}\theta^{3}\theta^{\pm}.\nonumber
\end{align}
The non-vanishing elements of the corresponding quantum metric are
\begin{equation}
g^{+-}=-q,\quad g^{33}=1,\quad g^{-+}=-q^{-1}.
\end{equation}
Covariant coordinates can be introduced by
\begin{equation}
\theta_{A}=g_{AB}\theta^{B},
\end{equation}
with $g_{AB}$ being the inverse of $g^{AB}$.

For antisymmetrized q-deformed Euclidean space with four dimensions we have
the relations
\begin{align}
(\theta^{i})^{2}  &  =0,\quad i=1,\ldots,4,\\
\theta^{1}\theta^{2}  &  =-q^{-1}\theta^{2}\theta^{1},\nonumber\\
\theta^{1}\theta^{3}  &  =-q^{-1}\theta^{3}\theta^{1},\nonumber\\
\theta^{2}\theta^{4}  &  =-q^{-1}\theta^{4}\theta^{2},\nonumber\\
\theta^{3}\theta^{4}  &  =-q^{-1}\theta^{3}\theta^{4},\nonumber\\
\theta^{1}\theta^{4}  &  =-\theta^{4}\theta^{1},\nonumber\\
\theta^{2}\theta^{3}  &  =-\theta^{3}\theta^{2}+\lambda\theta^{1}\theta
^{4},\nonumber
\end{align}
and its metric has the non-vanishing components
\begin{equation}
g^{14}=q^{-1},\quad g^{23}=g^{32}=1,\quad g^{41}=q.
\end{equation}

The generators of antisymmetrized q-deformed Minkowski space \cite{SWZ91,
Maj91, OSWZ92} are subject to the relations
\begin{align}
(\theta^{\mu})^{2}  &  =0,\quad\mu\in\{+,-,0\},\\
\theta^{3}\theta^{\pm}  &  =-q^{\mp2}\theta^{\pm}\theta^{3},\nonumber\\
\theta^{3}\theta^{3}  &  =\lambda\theta^{+}\theta^{-},\nonumber\\
\theta^{+}\theta^{-}  &  =-\theta^{-}\theta^{+},\nonumber\\
\theta^{\pm}\theta^{0}+\theta^{0}\theta^{\pm}  &  =\pm q^{\mp1}\lambda
\theta^{\pm}\theta^{3},\nonumber\\
\theta^{0}\theta^{3}+\theta^{3}\theta^{0}  &  =\lambda\theta^{+}\theta
^{-}.\nonumber
\end{align}
Instead of dealing with the coordinate $\theta^{3}$ or $\theta^{0}$ it is
often more convenient to work with the light-cone coordinate $\theta
^{3/0}=\theta^{3}-\theta^{0}$, for which we have the additional relations
\begin{align}
(\theta^{3/0})^{2}  &  =0,\\
\theta^{\pm}\theta^{3/0}  &  =-\theta^{3/0}\theta^{\pm},\nonumber\\
\theta^{0}\theta^{3/0}+\theta^{3/0}\theta^{0}  &  =-\lambda\theta^{+}%
\theta^{-},\nonumber\\
\theta^{\pm}\theta^{0}+q^{\pm2}\theta^{0}\theta^{\pm}  &  =\pm q^{\pm1}%
\lambda\theta^{\pm}\theta^{3/0},\nonumber\\
\theta^{3}\theta^{3/0}+\theta^{3/0}\theta^{3}  &  =-\lambda\theta^{+}%
\theta^{-}.\nonumber
\end{align}
Finally, we write down the non-vanishing entries of the matrix representing
q-deformed Minkowski metric:
\begin{equation}
\eta^{00}=-1,\quad\eta^{33}=1,\quad\eta^{+-}=-q,\quad\eta^{-+}=-q^{-1}.
\end{equation}

\end{document}